\documentclass[a4paper,12pt]{article}

\usepackage{amsmath,amssymb,mathtools}
\usepackage{color}
\usepackage{graphicx}
\usepackage{subfigure}
\usepackage{cite}
\usepackage{hyperref}
\usepackage{multirow,makecell}
\usepackage{textcomp}
\usepackage{wasysym}
\usepackage{setspace}
\usepackage{verbatim}
\usepackage{float}
\usepackage{ulem}
\usepackage[utf8]{inputenc}
\graphicspath{{figures/}}

\usepackage[text={17.2cm,25.2cm},centering]{geometry}

\numberwithin{equation}{section}

\begin{document}
\title{\textbf{$J/\Psi$ suppression in a rotating magnetized holographic QGP matter}}
	\author{Yan-Qing Zhao$^{1}$\footnote{zhaoyanqing@mails.ccnu.edu.cn } and Defu Hou$^{1}$\footnote{corresponding author: houdf@mail.ccnu.edu.cn }}

	\date{}

	\maketitle

	\vspace{-10mm}
	
	\begin{center}
		{\it
			$^{1}$ Institute of Particle Physics and Key Laboratory of Quark and Lepton Physics (MOS),
Central China Normal University, Wuhan 430079, China\\ \vspace{1mm}
		}
		\vspace{10mm}
	\end{center}

	\begin{abstract}
We study the dissociation effect of $J/\Psi$ in magnetized, rotating QGP matter  at finite temperature and chemical potential using gauge/gravity duality. By incorporating angular velocity into the holographic magnetic catalysis model, we analyze the influence of temperature, chemical potential, magnetic field, and angular velocity on the properties of $J/\Psi$ meson. The results reveal that temperature, chemical potential, and rotation enhance the dissociation effect and increase the effective mass in the QGP phase. However, the magnetic field suppresses dissociation, and its effect on the  effective mass is non-trivial. Additionally, we explore the interplay between magnetic field and rotation, identifying a critical angular velocity that determines the dominant effect. As a parallel study, we also examine the rotation effect in the holographic inverse magnetic catalysis model, although the magnetic field exhibits distinctly different behaviors in these two models, the impact of rotation on the dissociation effect of $J/\Psi$ is similar. Finally, we investigate the influence of electric field and demonstrate that it also speeds up the  $J/\Psi$ dissociation .
	\end{abstract}
	
	\baselineskip 18pt
	\thispagestyle{empty}
	\newpage
	
	\tableofcontents
\section{Introduction}\label{sec:01_intro}
In ultra-relativistic heavy-ion collisions, a new, strongly interacting matter state known as QGP~\cite{Collins:1974ky} is created. In order to fully understand the hot and dense plasma, as one of the best probes, the final yield of dilepton from heavy quarkonium ($J/\Psi$ and $\Upsilon(1S)$) is used to study the properties of QGP. The formation time of charm quark is about $\tau_c\sim 1/2m_c\approx0.06\, \text{fm/c}$ and about $\tau_b\sim 1/2m_b\approx0.02\, \text{fm/c}$ for bottom quark~\cite{Zhao:2021yfa}. The original concept of quarkonium suppression is from  Matsui and Satz~\cite{Matsui:1986dk}. The bound state is produced in the initial stage of  heavy-ion collision and then gets dissociated into free quarks when it passes through the QGP medium, which leads to the decrease of final output(the dilepton), this phenomenon is called quarkonium suppression. So far, there are two mechanisms that can be used to explain the reduction of quarkonium, namely: (1) Hot nuclear matter (HNM) effects{-----}the reduction is caused by the existence of QGP medium; The HNM effects include: \textcircled{1} The rotation of QGP medium, \textit{rotating effect};  \textcircled{2} The temperature for QGP medium, \textit{thermal effect}; \textcircled{3} The baryon chemical  potential, \textit{density effect}. (2) Cold nuclear matter (CNM) effects{-----}the decrease is resulted in by those matters other than QGP medium. The CNM effects include\cite{Andronic:2015wma, Zhang:2021snx}: \textcircled{1} Parton distribution in the nuclei, the \textit{nPDF effects}; \textcircled{2} Inelastic collision between the quarkonium and nucleons, the \textit{nuclear absorption};  \textcircled{3} Interaction of the quarkonium with co-moving particles, resulting in the melting of bound state, the \textit{co-mover dissociation}; \textcircled{4} The Parton energy loss during multiple scattering process, the \textit{energy loss}; \textcircled{5} Electromagnetic field generated by bystanders, the \textit{electromagnetic effect}.

Experimentally, the suppression of $J/\Psi$ production has already been observed in Au+Au collisions at $\sqrt{s_{NN}}=200\, \text{GeV}$  through the dimuon channel at STAR~\cite{STAR:2019fge}, where the $J/\Psi$ yields are measured in a large transverse momentum range($0.15\,\text{GeV/c}\leq p_T\leq12\,\text{GeV/c}$) from central to peripheral collisions. They find the $J/\Psi$ yields are suppressed by a factor of approximately $3$ for $p_T>5\,\text{GeV/c}$ in the $0-10\%$ most central collisions. Theoretically, heavy quarkonium may survive, due to the Coulomb attraction between quark and antiquark, as bound states above the deconfinement temperature. By using the maximum entropy method (MEM), Ref.~\cite{Asakawa:2003re} studies  the correlation functions of $J/\Psi$  at finite temperature on $32^3*(32-96)$  anisotropic lattices, the conclusions indicate $J/\Psi$ could  survive in the plasma up to $T_d\sim1.6 T_c$ and melt at $1.6T_c\leq T_d\leq 1.9T_c$. Lattice data ~\cite{Datta:2003ww} find $J/\Psi$ could  survive up to $1.5T_c$ and vanish at $3T_c$.

Gauge/gravity duality has emerged as a valuable tool for investigating the properties of heavy quarks. Holographic methods have been extensively employed to study heavy flavor properties~\cite{Zhao:2021ogc, Fujita:2009ca, Finazzo:2013rqy, Kim:2007rt, Gubser:2007zr, Finazzo:2014rca, Hashimoto:2014fha, Lee:2013oya, Fadafan:2013coa, BitaghsirFadafan:2015zjc, BitaghsirFadafan:2015yng, Braga:2020myi, Iatrakis:2015sua, Braga:2016oem, Bellantuono:2017msk, Feng:2019boe, Zollner:2020nnt, Zollner:2020cxb, Zhao:2019tjq, Wu:2012ns, Zhou:2021sdy, Zhang:2015gha, Kioumarsipour:2021zyg,Braga:2018zlu, Park:2009nb, Avila:2020ved, Hou:2007uk, MartinContreras:2021bis,Braga:2017bml}. In Ref.\cite{Zhao:2021ogc}, we investigate the vector meson spectral function using a dynamical AdS/QCD model. Our results reveal that the enhancement of heavy quarkonium dissociation is influenced by the magnetic field, chemical potential, and temperature. Specifically, we observe non-trivial changes in the peak position of $J/\Psi$, which we attribute to the interplay between the interaction of the two heavy quarks and the interaction of the medium with each heavy quark. Interestingly, by introducing a dilaton in the background action, we demonstrate that the magnetic field has a more pronounced impact on heavy meson dissociation when it aligns with the polarization, contrary to the findings in the EM model discussed in Ref.\cite{Braga:2018zlu}. In Ref.~\cite{Park:2009nb}, the binding energy of heavy quarkonium in the quark-gluon plasma (QGP) and hadronic phase is examined. The results reveal that the dissociation length of heavy mesons decreases with increasing temperature or quark chemical potential in the QGP. However, in the hadronic phase, the dissociation length increases with an increase in the chemical potential. This phenomenon is attributed to two distinct dissociation mechanisms: the screening of the interaction between heavy quarks by light quarks in the deconfinement phase, and the breaking of the heavy meson into two heavy-light quark bound states in the confinement phase. Furthermore, Ref.\cite{Avila:2020ved} identifies the existence of a second lower critical temperature for certain magnetic field intensities, below which stable mesons cease to exist. This is termed Magnetic Meson Melting (MMM), extending the understanding of meson melting in the presence of varying magnetic field intensities. Ref.\cite{Hou:2007uk} calculates the ratios of dissociation temperatures for $J/\Psi$ using the U-ansatz potential and finds agreement with lattice results within a factor of two. Additionally, Ref.~\cite{MartinContreras:2021bis} investigates heavy quarkonia spectroscopy at both zero and finite temperature using a bottom-up AdS/QCD approach, predicting the melting temperature of $J/\Psi$ to be approximately 415 MeV ($\sim2.92T_c$).

In this work, we investigate the behavior of heavy vector mesons in the rotating magnetized quark-gluon plasma (QGP) created at RHIC and LHC. As previously mentioned, although there are numerous non-perturbative methods available, including the state-of-the-art lattice QCD\cite{Chen:2021giw, Ding:2021ise, Ding:2012sp, Kaczmarek:2022ffn, Ding:2010yz}, for the study of heavy vector meson spectral functions, the emphasis has primarily been on exploring the thermal and density effects of the quark-gluon plasma (QGP) and the strong magnetic field effects. However, these properties merely capture a fraction of the complete QGP environment. Furthermore, the state-of-the-art lattice data not only support the scenario of inverse magnetic catalysis(IMC)\cite{Bornyakov:2013eya}, where the transition temperature decreases as the magnetic field increases for temperatures slightly above the critical temperature, but also present compelling evidence for magnetic catalysis(MC) \cite{DElia:2010abb} in the deconfinement phase diagram. In MC, the transition temperature increases as the magnetic field intensifies for temperatures not significantly higher than the critical temperature. Therefore, it is of utmost importance to take into account the MC, which holds an equally prominent position as inverse magnetic catalysis , along with other factors in order to simulate the influence of the QGP medium more accurately on the properties of heavy vector mesons. To bridge the existing gap, we also consider additional factors, such as medium rotation, electric fields, and magnetic field effects, which pose significant challenges for the current state-of-the-art lattice techniques. Using holographic methods, one of the non-perturbative approaches, we explore the interplay of these factors and their impact on heavy vector meson properties.

The rest structure of this paper is organized as follows. In section~\ref{sec:02}, we set up the holographic magnetic catalysis model. In section~\ref{sec:03}, we display the specific derivation process for calculating the spectral functions of $J/\Psi$ by introducing different field effects. In section~\ref{sec:04}, we show and discuss the meaningful results. Finally, this work is ended with a summary and a discussion in section~\ref{sec:05}.

\section{Holographic QCD model}\label{sec:02}

We consider a 5d EMD gravity system with a Maxwell field and a dilaton field as a thermal background for the corresponding hot, dense, magnetized QCD. The action is given as
\begin{eqnarray}\label{eq01}
  \mathcal{S} &=& \frac{1}{16\pi G_5}\int d^5 x \sqrt{-g}\left[ \mathcal{R}-\frac{1}{2}\nabla_\mu\phi_0\nabla^\mu\phi_0-\frac{f(\phi_0)}{4}F_{\mu\nu}F^{\mu\nu}-V(\phi_0)\right].
\end{eqnarray}
where $F_{\mu\nu}$ is the field strength tensor for U(1) gauge field, $f(\phi_0)$ is the gauge coupling kinetic function, $\phi_0$ is the dilaton field. $V(\phi_0)$ is the potential of the $\phi_0$.  As the dual system lives in a spatial plane, we choose Poincar\'{e} coordinates with $\tilde{z}$ the radial direction in the bulk. The metric ansatz reads~\cite{He:2020fdi}
\begin{align}\label{eq02}
 \tilde{g}_{\mu\nu}d\tilde{x}^\mu d\tilde{x}^\nu &= w_E(z)^2\bigg(-b(z)d\tilde{t}^2+\frac{d\tilde{z}^2}{b(z)}+(d\tilde{x}_2^2+d\tilde{x}_3^2)+e^{-Bz^2}d\tilde{x}_1^2\bigg), \notag \\
\phi_0 & =\phi_0(\tilde{z}),\quad A_{\mu}=(A_t(\tilde{z}),0,0,A_3(\tilde{x}_2),0),
\end{align}
with
\begin{align}\label{eqa1}
  b(\tilde{z})&=1-\frac{I_1(\tilde{z})}{I_1(\tilde{z}_h)}+\frac{\mu^2}{I_2^2(\tilde{z}_h)I_1(\tilde{z}_h)}(I_1(\tilde{z}_h)I_3(\tilde{z})-I_1(\tilde{z})I_3(\tilde{z}_h))+\frac{B^2}{I_1(\tilde{z}_h)}(I_1(\tilde{z}_h)I_4(\tilde{z})-I_1(\tilde{z})I_4(\tilde{z}_h)),\notag\\
  I_1(\tilde{z})&=\int_0^{\tilde{z}}\frac{dy}{w_E^3e^{-\frac{1}{2}By^2}},\quad\quad\quad
  I_2(\tilde{z})=\int_0^{\tilde{z}} \frac{dy}{w_Efe^{-\frac{1}{2}By^2}},\quad\quad\quad
  I_3(\tilde{z})=\int_0^{\tilde{z}} I_1'(y)I_2(y)dy,\notag\\
  I_4(\tilde{z})&=\int_0^{\tilde{z}} I_1'(y)I_5(y)dy,\quad\quad
  I_5(\tilde{z})=\int_0^{\tilde{z}} w_Efe^{-\frac{1}{2}By^2}dy,
\end{align}
where $w_E(\tilde{z})=\frac{1}{\tilde{z}}e^{\frac{-c\tilde{z}^2}{3}-p\tilde{z}^4}$ denotes the warped factor, $c=1.16$,  $p=0.273$ determines the transition point at $\tilde{\mu}=B=0$ fixed by fitting the lattice QCD data~\cite{He:2013qq,Li:2017tdz}. It should be noted that we use the notation, $\tilde{x}^\mu=({\tilde{t},\tilde{z},\tilde{x}_1,\tilde{x}_2,\tilde{x}_3})$, to represent the static frame and $x^\mu=({t,z,x_1, x_2,x_3})$ to represent the rotating frame. The AdS boundary at $\tilde{z}=0$. Here we have turned on a constant magnetic field $B$ along the $\tilde{x}_1$ direction in the dual field theory.

The Hawking temperature can be calculated by surface gravity
\begin{equation}\label{eqa2}
  T(\tilde{z}_h,\tilde{\mu},B)=\frac{I_1'(\tilde{z}_h)}{4\pi I_1(\tilde{z}_h)}(1-\tilde{\mu}^2\frac{I_1(\tilde{z}_h)I_2(\tilde{z}_h)-I_3(\tilde{z}_h)}{I_2^2(\tilde{z}_h)}-B^2(I_1(\tilde{z}_h)I_5(\tilde{z}_h)-I_4(\tilde{z}_h))).
\end{equation}
In this paper, the deconfinement phase transition temperature for zero chemical potential and magnetic field is at $\tilde{T}_c=0.6\text{GeV}$~\cite{He:2020fdi}.
\section{The spectral functions}\label{sec:03}

In this section, we will calculate the spectral function for $J/\Psi$ state by a phenomenological model proposed in Ref.~\cite{Braga:2017bml}. In order to go smoothly later, we assume the metric has general forms as follows,
\begin{equation}\label{A1}
  ds^2=-g_{tt} dt^2+g_{x_1x_1}dx_1^2+g_{tx_1}dtdx_1+g_{x_1t}dx_1dt+g_{x_{2,3}x_{2,3}}dx_{2,3}^2+g_{zz} dz^2.
\end{equation}
The vector field $A_m=(A_\mu, A_z)(\mu=0,1,2,3)$ is used to represent the heavy quarkonium, which is dual to the gauge theory current $J^\mu=\overline{\Psi}\gamma^\mu\Psi$. The standard Maxwell action takes the following form
\begin{equation}\label{eq06}
   S=-\int d^4xdz \frac{Q}{4} F_{mn}F^{mn},
\end{equation}
where $F_{mn}=\partial_mA_n-\partial_nA_m$, $Q=\frac{\sqrt{-g}}{h(\phi){g_5}^2}$, $h(\phi)=e^{\phi(z)}$. The function $\phi(z)$ is used to parameterize vector mesons,
\begin{equation}\label{eq07}
  \phi(z)=\kappa^2z^2+Mz+\tanh(\frac{1}{Mz}-\frac{\kappa}{\sqrt{\Gamma}}).
\end{equation}
where $\kappa$ labels the quark mass, $\Gamma$ is the string tension of the quark pair and $M$ denotes a large mass related to the heavy quarkonium non-hadronic decay. The value of three energy parameters for charmonium in the scalar field, determined by fitting the spectrum of masses~\cite{Braga:2018zlu}, are respectively:
\begin{align}\label{eq08}
  \kappa_c & =1.2\text{GeV},\quad \sqrt{\Gamma_c}=0.55\text{GeV},\quad M_c=2.2\text{GeV}\,.
\end{align}

The spectral functions for $J/\Psi$ state will be calculated with the help of the membrane paradigm~\cite{Iqbal:2008by}. The equation of motion obtained from Eq.~\eqref{eq06} are as follows
\begin{equation}\label{AA5}
 \partial_m(QF^{mn})=\partial_z(QF^{zn})+\partial_\mu(QF^{\mu n}),
\end{equation}
where $F^{mn}=g^{m\alpha}g^{n\beta}F_{\alpha\beta}$, $n=(0,1,2,3,4)$ and $\mu=(0,1,2,3)$.  For the $z$-foliation, the conjugate momentum of the gauge field $A^\mu$ is given by the following formula:
\begin{equation}\label{eq10}
 j^\mu=-QF^{z\mu}.
\end{equation}
Supposing plane wave solution for vector field $A^\mu$ propagates in the $x_1$ direction. The equation of motion~\eqref{AA5} can be written as two parts: longitudinal-the fluctuations along $(t,x_1)$; transverse-fluctuations along $(x_2,x_3)$. Combined with Eq.\eqref{eq10}, the dynamical equations for longitudinal case from components $t, x_1$ and $z$ of Eq.\eqref{AA5} can be expressed as
 \begin{align}
 -\partial_zj^{t}-Q(g^{x_1x_1}g^{tt}+g^{x_1t}g^{tx_1})\partial_{x_1}F_{x_1t} & =0, \label{eq11}\\
  -\partial_zj^{x_1}+Q(g^{tt}g^{x_1x_1}+g^{tx_1}g^{x_1t})\partial_tF_{x_1 t} & =0,\label{eq12}
\end{align}
The flow $j^\mu$ conservation equation and the Bianchi identity can be written as:
\begin{align}
 \partial_{x_1}j^{x_1}+\partial_tj^t & =0\label{eq13},\\
 \partial_zF_{x_1t}-\frac{g_{zz}}{Q}\partial_t[g_{x_1x_1}j^{x_1}+g_{x_1t}j^{t}]-\frac{g_{zz}}{Q}\partial_{x_1}[g_{tt}j^{t}-g_{tx_1}j^{x_1}]&=0.\label{eq14}
\end{align}
The longitudinal "conductivity" and its derivative are defined as
\begin{align}
  \sigma_L(\omega,z) & =\frac{j^{x_1}(\omega,z)}{F_{x_1t}(\omega,z)},\label{eq15} \\
  \partial_z \sigma_L(\omega,z) & =\frac{\partial_zj^{x_1}}{F_{x_1t}}-\frac{j^{x_1}}{F_{x_1t}^2}\partial_zF_{x_1t}\label{eq16}.
\end{align}
Kubo's formula shows that the five-dimensional "conductivity" at the boundary is related to the retarded Green's function:
\begin{equation}\label{eq17}
  \sigma_L(\omega)=\frac{-G_R^L(\omega)}{i\omega}.
\end{equation}
where $\sigma_L$ is interpreted as the longitudinal AC conductivity. In order to obtain flow equation ~\eqref{eq16}, we assume $A_\mu=A_n(p,z)e^{-i\omega t+ipx_1}$, where $A_n(p,z)$ is the quasinormal modes. Therefore, we have $\partial_tF_{x_1t}=-i\omega F_{x_1t}$, $\partial_tj^{x_1}=-i\omega j^{x_1}$. Finally, by using Eq.\eqref{eq12}, Eq.\eqref{eq13}, Eq.\eqref{eq14} and taking the momentum limit $P=(\omega,0,0,0)$, the Eq.\eqref{eq16} can be written as
\begin{equation}\label{eq18}
 \partial_z \sigma_L(\omega,z)=\frac{i\omega g_{x_1x_1}g_{zz}}{Q}(\sigma_L^2-\frac{Q^2(g^{tt}g^{x_1x_1}+g^{tx_1}g^{x_1t})}{ g_{x_1x_1}g_{zz}}).
\end{equation}
The initial condition for solving the equation can be obtained by  requiring regularity at the horizon $\partial_z \sigma_L(\omega,z)=0$. The dynamical equation of transverse channel is as follows:
\begin{align}
\partial_zj^{x_2}\!-Q[g^{tx_1}g^{x_2x_2}\partial_{t}F_{x_1x_2}\!-g^{tt}g^{x_2x_2}\partial_{t}F_{tx_2}\!+
g^{x_1t}g^{x_2x_2}\partial_{x_1}F_{tx_2}\!+g^{x_1x_1}g^{x_2x_2}\partial_{x_1}F_{x_1x_2}] & = 0,\label{eq19}\\
\frac{g_{x_2x_2}g_{zz}}{Q}\partial_tj^{x_2}+\partial_{z}F_{t x_2} & = 0,\label{eqa19}\\
\partial_{x_1}F_{tx_2}+\partial_tF_{x_2x_1} & = 0.\label{eqaa19}
\end{align}
The transverse "conductivity" and its derivative are defined as
\begin{align}
  \sigma_T(\omega,z) & =\frac{j^{x_2}(\omega,\overrightarrow{p},z)}{F_{x_2t}(\omega,\overrightarrow{p},z)},\label{eqa20} \\
  \partial_z \sigma_T(\omega,z) & =\frac{\partial_zj^{x_2}}{F_{x_2t}}-\frac{j^{x_2}}{F_{x_2t}^2}\partial_zF_{x_2t}\label{eqaa20}.
\end{align}
Similarly,  we have $\partial_{t}F_{x_1x_2}=-i\omega F_{x_1x_2}$,\,$\partial_tF_{tx_2}=-i\omega F_{tx_2}$,\,$\partial_tj^{x_2}=-i\omega j^{x_2}$. Then the transverse flow equation \eqref{eqaa20} can be written as
\begin{align}\label{eq3a20}
\begin{split}
  \partial_z \sigma_T(\omega,z)
  &=\frac{i\omega g_{zz}g_{x_2x_2}}{Q}(\sigma_T^2-\frac{Q^2g^{zz}g^{tt}}{g_{x_2x_2}^2}).
\end{split}
\end{align}

It is not difficult to find that the metric Eq.\eqref{eq02} restores the SO(3) invariance and the flow equations Eq.\eqref{eq18} and Eq.\eqref{eq3a20} have the same form when magnetic field $B=0 \text{GeV}^2$.
The spectral function is defined by the retarded Green's function
\begin{equation}\label{eq20}
  \rho(\omega)\equiv-Im G_R(\omega)=\omega Re\,\sigma(\omega,0)
\end{equation}
%

\subsection{Turning on the angular momentum}\label{sec:3.1}

In the early stage of non-central heavy-ion collisions,  produced partons have a large initial orbital angular momentum $J\propto b\sqrt{s_{NN}}$ where $b$ is the impact parameter and $\sqrt{s_{NN}}$ the nucleon-nucleon center-of-mass energy. Although, at the stage of initial impact,  most of the angular momentum is carried away by the so-called "spectators", there is a considerable part that remains in the produced QGP~\cite{Jiang:2016woz}. Star collaboration find, by studying the global $\Lambda$ polarization in nuclear collisions, that the average vorticity of QGP could reach $\Omega\sim10^{21}/s$~\cite{STAR:2017ckg}.

Following~\cite{BravoGaete:2017dso, Nadi:2019bqu, Erices:2017izj, Chen:2020ath, Zhou:2021sdy}, we extend the holographic magnetic catalysis model to the situation of rotation with a planar horizon. For a general metric in the rest frame
\begin{equation}\label{eq21}
  d\tilde{s}^2=-\tilde{g}_{tt}d\tilde{t}^2+\tilde{g}_{zz}d\tilde{z}^2+\tilde{g}_{x_1x_1}d\tilde{x}_1^2+\tilde{g}_{x_{2,3}x_{2,3}}d\tilde{x}_{2,3}^2,
\end{equation}
to introduce the rotation effect, it is convenient to  split the 3-dimensional space into two parts as $\mathcal{M}_3=\mathbb{R}\times \Sigma_2$. Then we have
\begin{equation}\label{eq21p}
  d\tilde{s}^2=-\tilde{g}_{tt}d\tilde{t}^2+\tilde{g}_{zz}d\tilde{z}^2+\tilde{g}_{x_1x_1}\l^2d\tilde{\theta}^2+\tilde{g}_{x_{2,3}x_{2,3}}d\sigma^2,
\end{equation}
where $l$ denotes the fixed distance to rotating axis and $d \sigma^2$ represents the line element of $\Sigma_2$. Then the angular momentum will be turned on in the $l\tilde{\theta}$ direction through the standard Lorentz transformation,
\begin{equation}\label{eq22}
  \tilde{t}\rightarrow\gamma(t+\Omega l^2\theta), \quad \quad \tilde{\theta}\rightarrow\gamma(\theta+\Omega t),
\end{equation}
 where $\gamma=\frac{1}{\sqrt{1-\Omega^2 l^2}}$ is the usual Lorentz factor. It is estimated that the size of QGP may be around 4-8fm (RHIC) and 6-11fm (LHC)~\cite{Zhang:2001vk}. Without loss of generality, we use $l=1\,\text{GeV}^{-1}$ in the subsequent numerical calculation.   The metric \eqref{eq21} changes to the following form,
\begin{equation}\label{eq23}
  ds^2=\gamma^2(\tilde{g}_{x_1x_1}\Omega^2 l^2-\tilde{g}_{tt})dt^2+2\gamma^2\Omega l^2 (\tilde{g}_{x_1x_1}-\tilde{g}_{tt} )dtd\theta+\gamma^2 (\tilde{g}_{x_1x_1}-\Omega^2 l^2\tilde{g}_{tt}) l^2 d\theta^2+\tilde{g}_{zz}dz^2+\tilde{g}_{x_{2,3}x_{2,3}}d\sigma^2.
\end{equation}
Then the Hawking temperature and chemical potential of the rotating black hole can be calculated by
\begin{align}\label{eq26}
 T(z_h,\mu,B,\Omega)&=\tilde{T}(\tilde{z}_h,\tilde{\mu},B)\sqrt{1-\Omega^2l^2},\notag\\
 \mu(\Omega)&=\tilde{\mu}\sqrt{1-\Omega^2l^2}.
\end{align}
Next, we calculate the spectral function of heavy quarkonium for rotating case.  Suppose that, in the limit of zero momentum, the plane wave solution of the vector field has the form $A_\mu(t,z)=e^{-i \omega t} A_\mu(z,\omega)$.  Due to the appearance of rotation and magnetic field destroying the rotational symmetry of space,  the EOM\eqref{AA5} can be written in two varying channels: longitudinal-the direction parallel to the magnetic field; transverse-the direction perpendicular to the magnetic field. With the help of Eq.\eqref{eq18} and Eq.\eqref{eq3a20}, the flow equation can be written as
\begin{align}
 \partial_z \sigma_L(\omega,z)=i\omega \Xi^{//}(\sigma_L(\omega,z)^2-(\Delta^{//})^2),\notag\\
 \partial_z \sigma_T(\omega,z)=i\omega \Xi^{\perp}(\sigma_T(\omega,z)^2-(\Delta^{\perp})^2)\label{eq28}.
\end{align}
(1) When the direction of angular velocity is parallel to the direction of the magnetic field,
    \begin{align}\label{eq29}
      \Xi^{//}&=\frac{ e^{\frac{-B z^2}{2}+\phi (z)} \left(l^2 \Omega ^2 e^{B z^2}
   b(z)-1\right)}{b(z) \left(l^2 \Omega ^2-1\right) w_E(z)}  ,\nonumber\\
   \Delta^{//}&=w_E(z) e^{\frac{B z^2}{2}- \phi(z)}\sqrt{\frac{l^2 \Omega ^2-1}{l^2 \Omega ^2 e^{B z^2} b(z)-1}},\nonumber\quad\quad\quad(\Omega//B//P)\\
   \Xi^{\perp}&=\frac{e^{\frac{B z^2}{2}+\phi (z)}}{b(z) w_E(z)},\nonumber\\
     \Delta^{\perp}&=w_E(z) e^{-\frac{B z^2}{2}-\phi (z)}\sqrt{\frac{ l^2 \Omega ^2 e^{Bz^2} b(z)-1}{l^2 \Omega ^2-1}}.\,\,\,\,\quad\quad(\Omega//B\perp P)
    \end{align}
(2)When the direction of angular velocity is perpendicular to the direction of the magnetic field,
    \begin{align}\label{eq30}
     \Xi^{//}&=\frac{e^{B z^2+\phi (z)} \left(1-l^2 \Omega ^2 b(z)\right)}{b(z) w_E(z) \sqrt{e^{B z^2}
   \left(l^4 \Omega ^4-l^2 \Omega ^2+1\right)-l^2 \Omega ^2+\frac{l^2
   \Omega ^2 \left(1-e^{B z^2}\right)}{b(z)}}},\notag\\
    \Delta^{//}&=w_E(z) e^{-\frac{B z^2}{2}-\phi (z)} \sqrt{\frac{l^2 \Omega
   ^2-1}{l^2 \Omega ^2 g(z)-1}},\nonumber \,\quad\quad\quad\quad\quad\quad\quad(\Omega//P\perp B)\\
   \Xi^{\perp}&=\frac{\left(1-l^2 \Omega ^2\right) e^{\phi (z)}}{w_E(z) \sqrt{b(z)
   \left(b(z) \left(e^{B z^2} \left(l^4 \Omega ^4-l^2 \Omega
   ^2+1\right)-l^2 \Omega ^2\right)+l^2 \Omega ^2 \left(1-e^{B
   z^2}\right)\right)}},\notag\\
    \Delta^{\perp}&=w_E(z) e^{\frac{B z^2}{2}-\phi (z)} \sqrt{\frac{l^2 \Omega ^2
   g(z)-1}{l^2 \Omega ^2-1}},\nonumber \,\,\,\,\quad\quad\quad\quad\quad\quad\quad(\Omega\perp B//P)\\
   \Xi^{\perp}&=\frac{\left(1-l^2 \Omega ^2\right) e^{B z^2+\phi (z)}}{w_E(z)
   \sqrt{b(z) \left(b(z) \left(e^{B z^2} \left(l^4 \Omega ^4-l^2 \Omega
   ^2+1\right)-l^2 \Omega ^2\right)+l^2 \Omega ^2 \left(1-e^{B
   z^2}\right)\right)}},\notag\\
    \Delta^{\perp}&=w_E(z) e^{-\frac{B z^2}{2}-\phi (z)} \sqrt{\frac{l^2 \Omega ^2
   g(z)-1}{l^2 \Omega ^2-1}} . \,\,\quad\quad\quad\quad\quad\quad\quad(\Omega\perp B\perp P)
    \end{align}
For vanishing angular momentum $\Omega$ and magnetic field $B$, one can easily check that the Eq.\eqref{eq29}-\eqref{eq30} have the same forms:
\begin{align}\label{eq34}
     \Xi^{//}=\Xi^{\perp}=\frac{ e^{\phi (z)} }{b(z) w_E(z)}  ,\quad \Delta^{//}=\Delta^{\perp}=w_E(z) e^{-\phi(z)}.
   \end{align}
%

\subsection{Adding a constant electric field to the background}

In this subsection, a constant electric field is added on the D-brane, see Ref.\cite{Matsuo:2006ws} for more information. The field strength tensor  can be expressed as $F=Edt\wedge dx_1$ where $E$ is the electric field along the $x_1$ direction. Since the equation of motion only depends on the field strength tensor, this ansatz is still a good solution to supergravity and  is the minimal setup to study the E-field correction for the corresponding field theory. Then one can write the E-field metric as

\begin{equation}\label{eq35}
 \mathcal{ F}_{\mu\nu}=
 \left (
 \begin{matrix}
   0 & 2\pi\alpha'E & 0&0 \\
   -2\pi\alpha'E &0 & 0&0 \\
   0 & 0& 0&0\\
   0 & 0& 0&0
  \end{matrix}
  \right ).
\end{equation}

Further, the background metric can be written as $ds^2=G_{\mu\nu}+\mathcal{F}_{\mu\nu}$ where $G_{\mu\nu}$ is from Eq.\eqref{eq21}.  The flow equation Eq.\eqref{eq28} becomes:

(1) When the direction of electric field is parallel to the direction of magnetic field,
    \begin{align}\label{eq36}
      \Xi^{//}&=\frac{e^{\phi (z)-B z^2}}{b(z) w_E(z) \sqrt{e^{-B z^2}-\frac{4 \pi
   ^2 \alpha ^2 E^2 w_E(z)^6}{b(z)}}} ,\quad \Delta^{//}=w_E(z) e^{\frac{B z^2}{2}-\phi (z)},\quad(E//B//P)   \notag\\
   \Xi^{\perp}&=\frac{w_E(z) e^{\phi (z)}}{b(z) \sqrt{e^{-B z^2}
  w_E(z)^4-\frac{4 \pi ^2 \alpha ^2 E^2}{b(z)}}},\quad \Delta^{\perp}=w_E(z) e^{-\frac{B z^2}{2}-\phi (z)}.\,\,\,\,\quad\quad(E//B\perp P)
    \end{align}
(2)When the direction of electric field is perpendicular to the direction of magnetic field,
    \begin{align}\label{eq38}
     \Xi^{//}&=\frac{w_E(z) e^{\frac{B z^2}{2}+\phi (z)}}{b(z) \sqrt{\frac{b(z)
   w_E(z)^4-4 \pi ^2 \alpha ^2 E^2}{b(z)}}},\quad \Delta^{//}=w_E(z) e^{-\frac{B z^2}{2}-\phi (z)},\,\quad(E//P\perp B)\notag\\
   \Xi^{\perp}&=\frac{w_E(z) e^{\phi (z)-\frac{B z^2}{2}}}{b(z) \sqrt{\frac{b(z)
   w_E(z)^4-4 \pi ^2 \alpha ^2 E^2}{b(z)}}},\quad \Delta^{\perp}=w_E(z) e^{\frac{B z^2}{2}-\phi (z)},\,\,\,\,\,\quad(E\perp B//P)\notag\\
   \Xi^{\perp}&=\frac{w_E(z) e^{\frac{B z^2}{2}+\phi (z)}}{b(z) \sqrt{\frac{b(z)
   w_E(z)^4-4 \pi ^2 \alpha ^2 E^2}{b(z)}}},\quad \Delta^{\perp}=w_E(z) e^{-\frac{B z^2}{2}-\phi (z)} .\,\,\,\quad(E\perp B\perp P)
    \end{align}
For vanishing E-field and B-field, one can find that the Eq.\eqref{eq36}-\eqref{eq38} have the same forms as Eq.\eqref{eq34}. In addition, one can find when the E-field is perpendicular to magnetic field, the flow equation has the same form for $E//P\perp B$ and $E\perp B\perp P$.

\section{ Numerical Results of spectral  functions in a magnetized rotating plasma}\label{sec:04}
%
\begin{figure}
  \centering
  \includegraphics[width=0.49\textwidth]{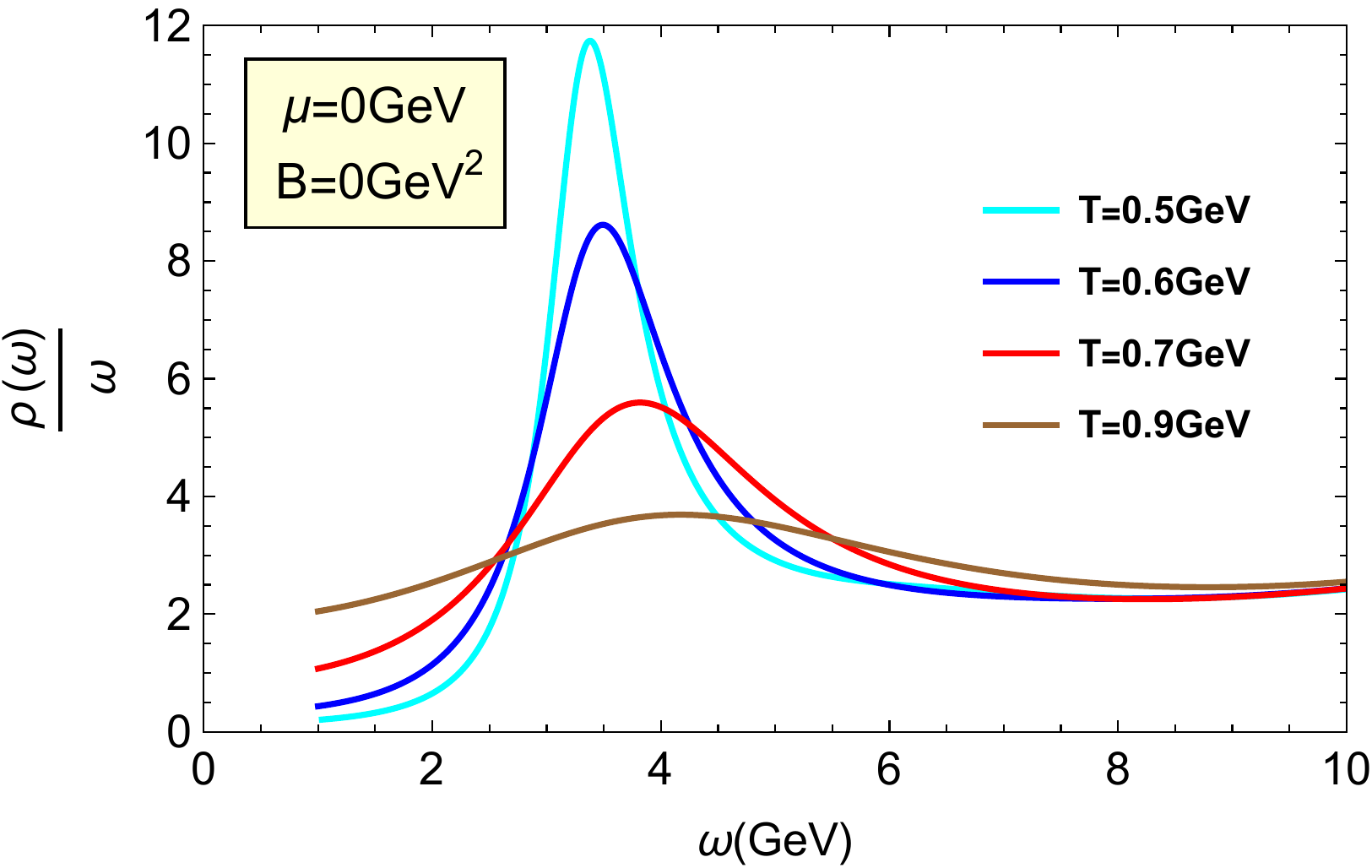}
  \includegraphics[width=0.49\textwidth]{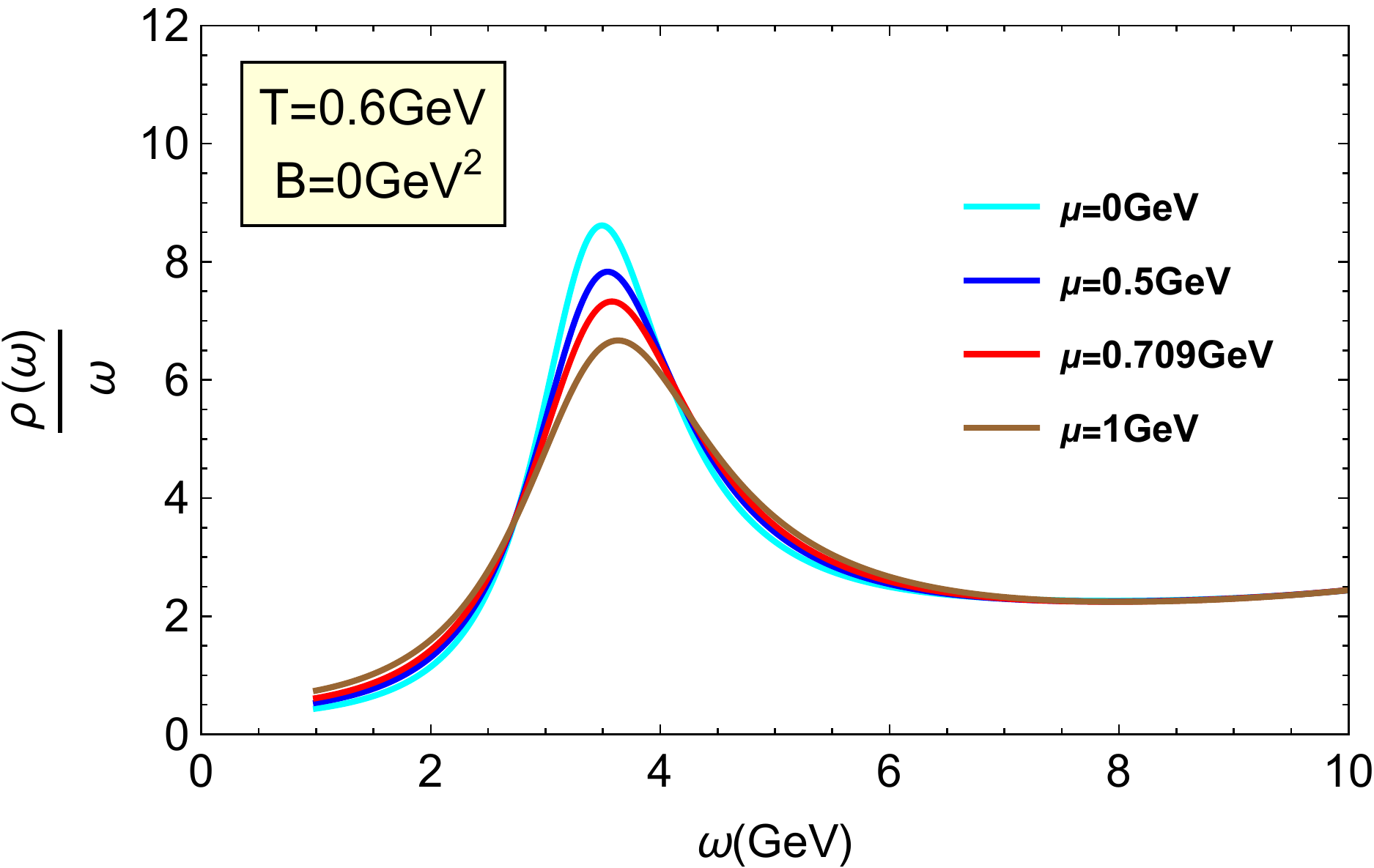}\\
  \caption{The spectral functions of the $J/\Psi$ state with different temperature $T$ (left panel) at $\mu=0\, \text{GeV}$ and $B=0\, \text{GeV}^2$ and different chemical potential $\mu$ (right panel) at $B=0\, \text{GeV}^2$ and $T=0.6\, \text{GeV}$ for magnetic catalysis model. From top to bottom, the curves represent $T=0.5, 0.6, 0.7, 0.9\, \text{GeV}$ in the left panel respectively, and  those denote $\mu=0, 0.5, 0.709, 1 \,\text{GeV}$ in the right panel respectively. }\label{Fig1}
\end{figure}
\begin{figure}
  \centering
  \includegraphics[width=0.6\textwidth]{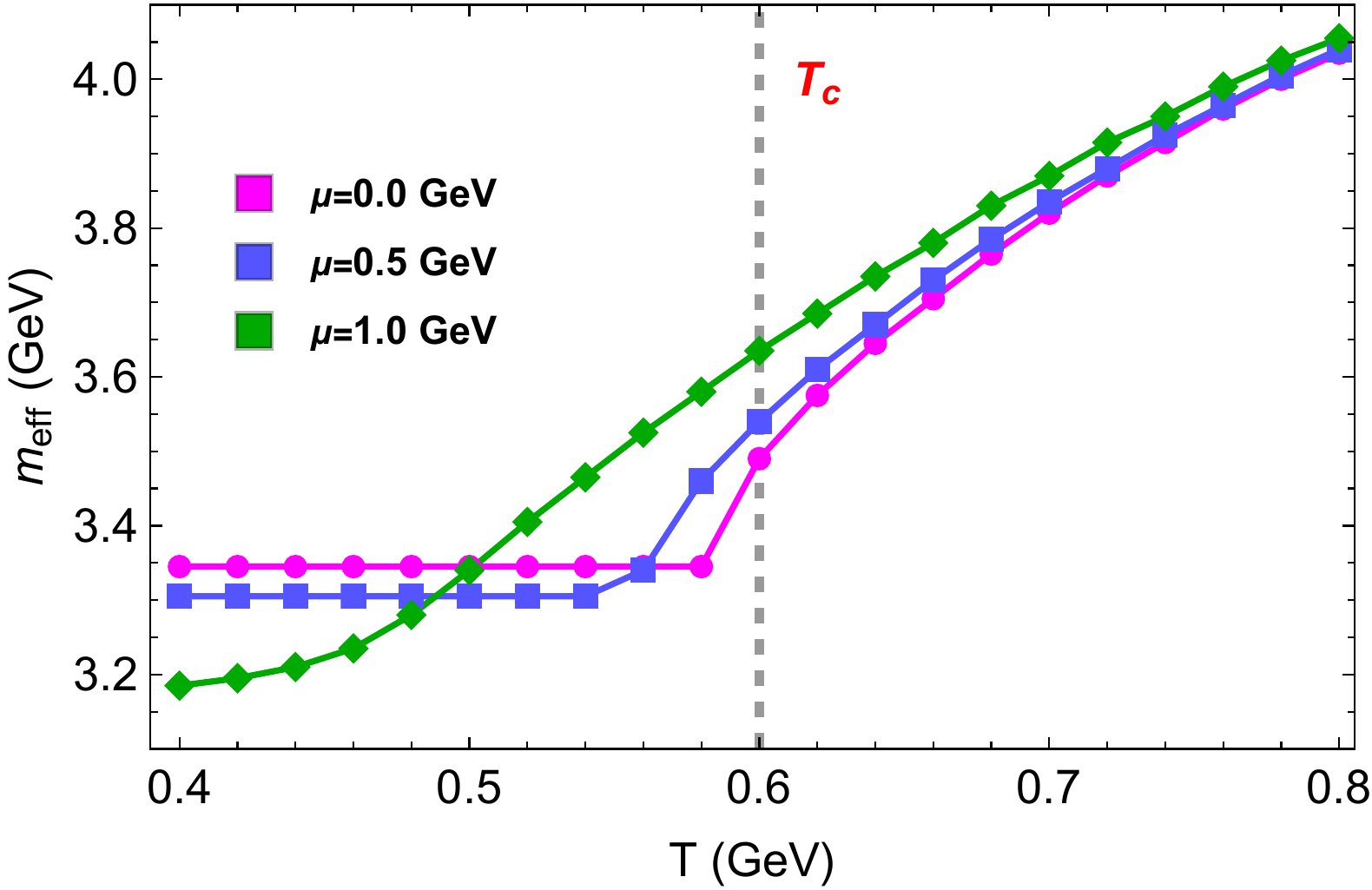}\\
  \caption{The effective mass of $J/\Psi$ state as a function of temperature in different chemical potentials for magnetic catalysis model.}\label{Figp1}
\end{figure}
The spectral function is described by Eq.\eqref{eq20}. Firstly, we plot our numerical results of the spectral function for the different temperatures and chemical potential in Fig.\ref{Fig1}. One can find the increase of temperature and chemical potential decrease the height and increase the width of spectral function peak. The decrease in peak height and the increase in peak width represent the enhancement of dissociation effect for heavy quarkonium.  So we can get that increasing temperature and chemical potential promote the dissociation effect of bound state. In Fig.\ref{Figp1}, we display the effective mass corresponding to the location of spectral function peak as a function of temperature in varying chemical potentials. As increasing temperature, the effective mass remains unchanged in the lower temperature regime, while that increases in the higher temperature regime. In the lower temperature regime, chemical potential reduces the effective mass. In the higher temperature regime, chemical potential increases the effective mass. It is easy to understand this phenomenon that increasing temperature enlarges the distance of quark anti-quark pair, which leads to the interaction between quark/anti-quark and medium becoming stronger. Therefore, the effective mass is larger with the increase of temperature.

\subsection{Turning on a constant magnetic field}
%
\begin{figure}[t]
  \centering
     \includegraphics[width=0.49\textwidth]{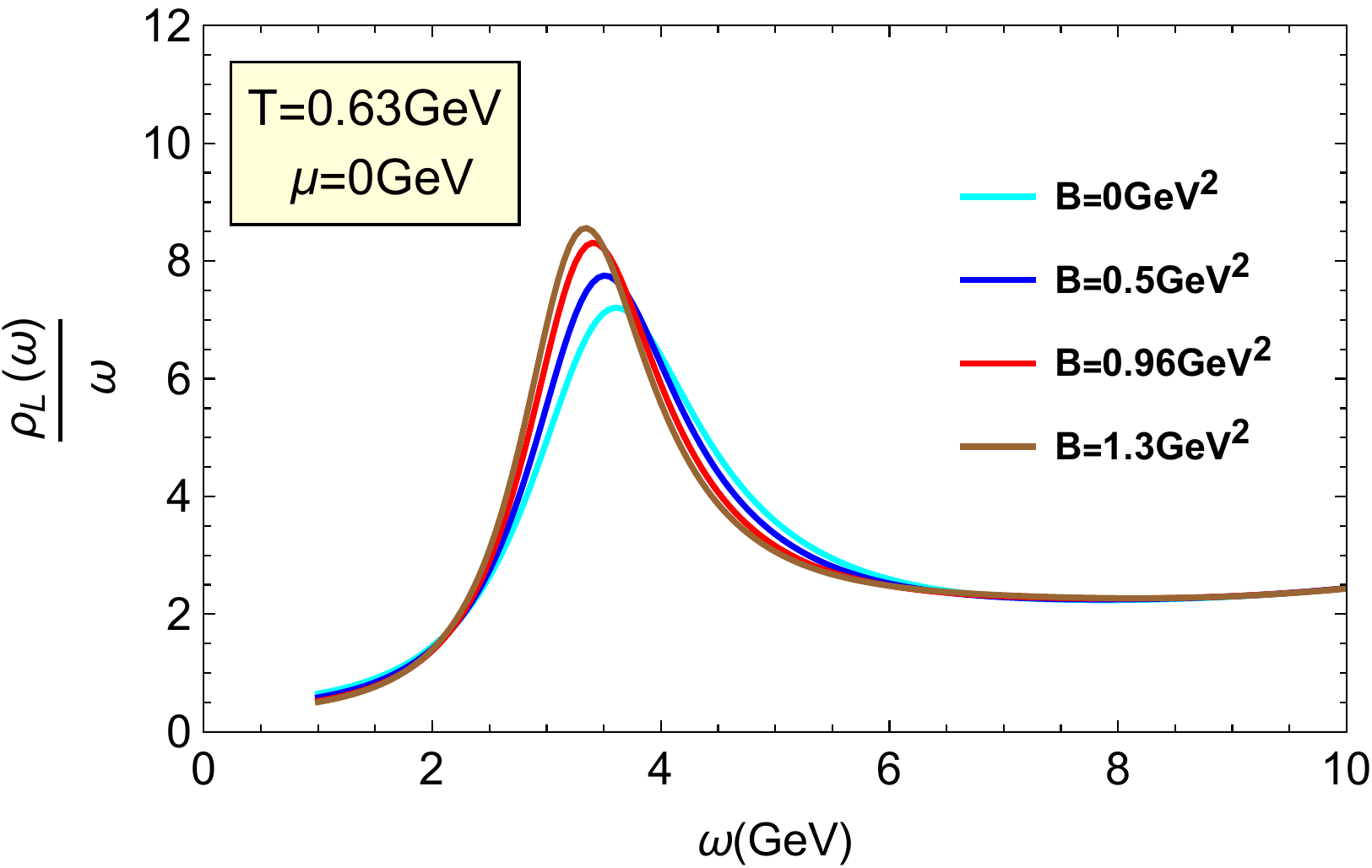}
     \includegraphics[width=0.49\textwidth]{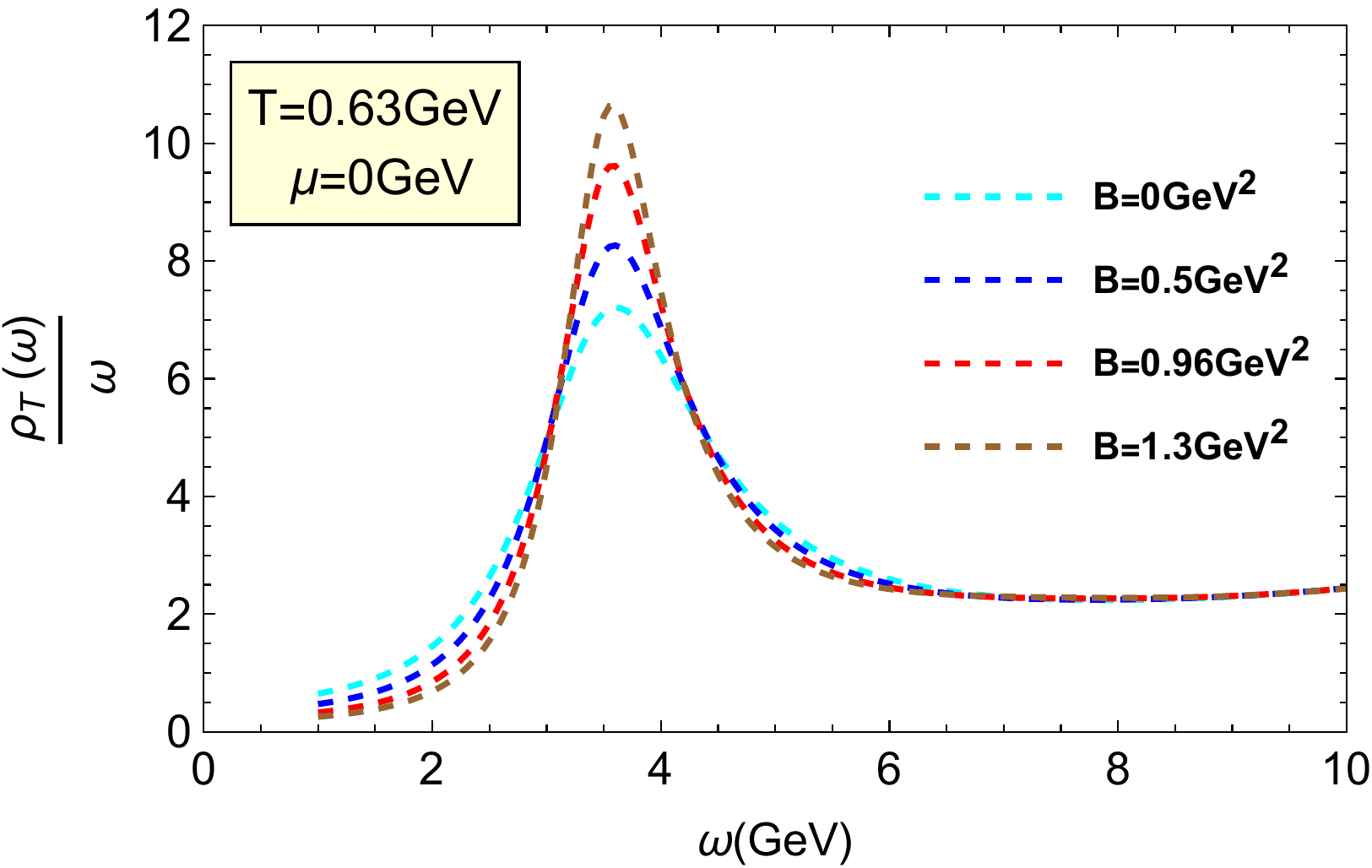}\\
  \caption{The spectral functions of the $J/\Psi$ state with different magnetic field $B$ at $\mu=0\,\text{GeV}$ and $T=0.63\,\text{GeV}$ for magnetic catalysis model. From bottom to top, the curves represent $B=0, 0.5, 0.96, 1.3\, \text{GeV}^2$ respectively. }\label{Fig3}
\end{figure}
\begin{figure}[t]
  \centering
     \includegraphics[width=0.49\textwidth]{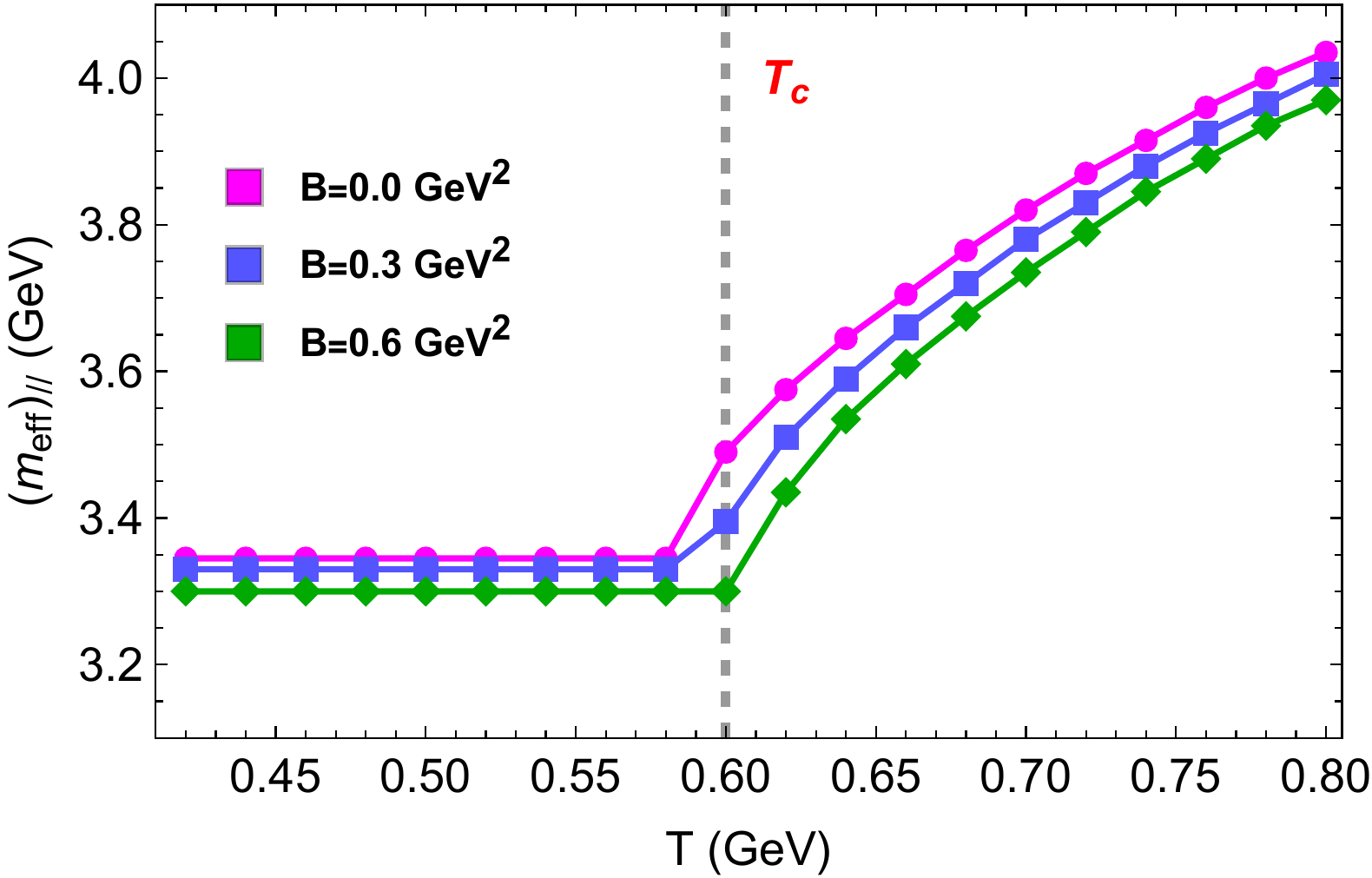}
     \includegraphics[width=0.49\textwidth]{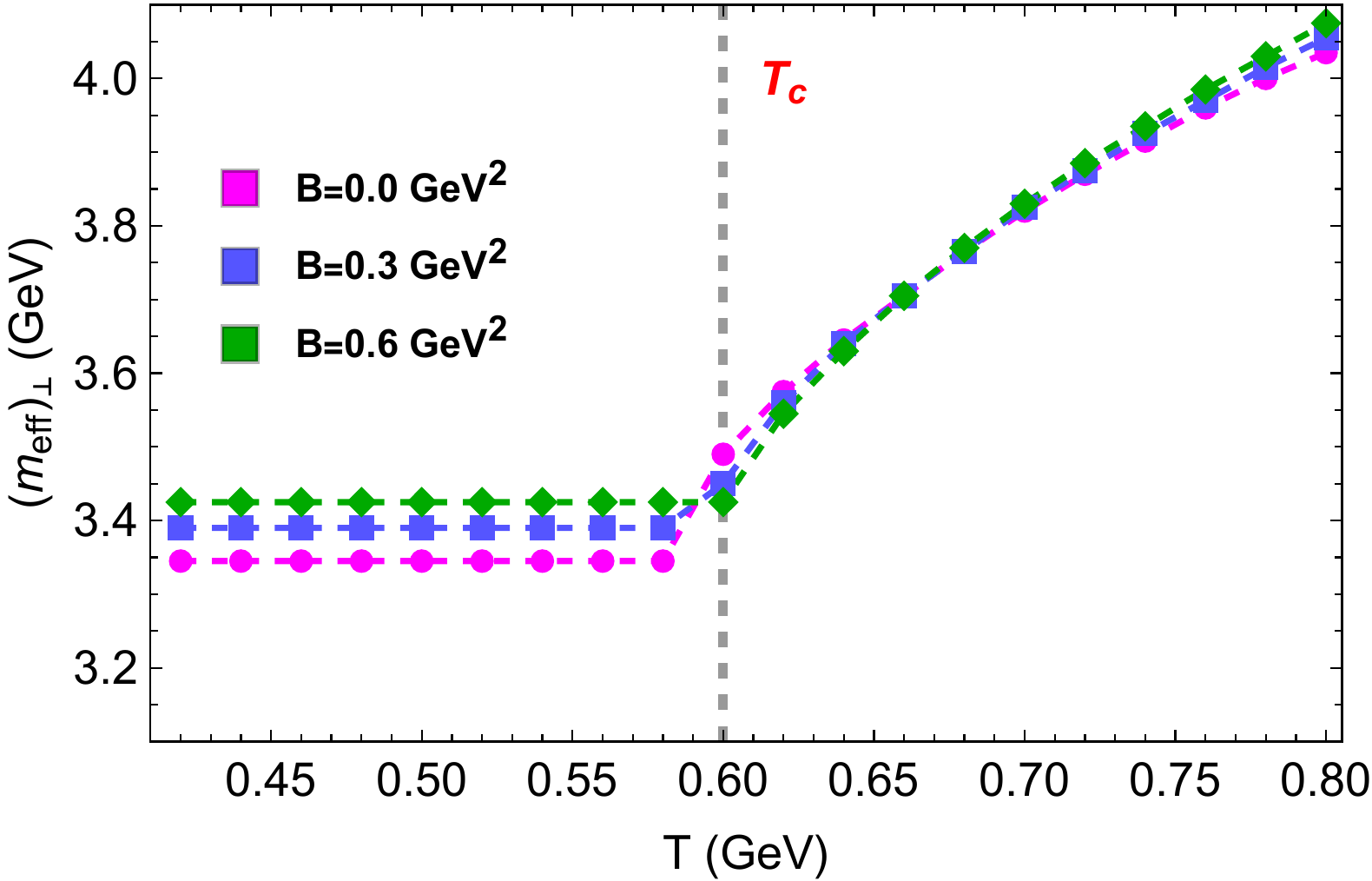}\\
  \caption{The effective mass of $J/\Psi$ state as a function of temperature in different magnetic fields for magnetic catalysis model. The left picture is for magnetic field parallel to polarization and the right one is for magnetic field perpendicular to polarization. }\label{Figp2}
\end{figure}

The spectral function with respect to the magnetic field is presented in Fig.\ref{Fig3} at $\mu=0\,\text{GeV}, T=0.63\,\text{GeV}$. The left panel is for magnetic field parallel to polarization and the right one is for magnetic field perpendicular to polarization.  A phenomenon completely opposite to the chemical potential effect and temperature effect is observed. Whether the magnetic field is parallel or perpendicular to the polarization, the strengthened magnetic field increases the peak height and reduces the peak width. That means the presence of magnetic field suppresses the dissociation of bound state, which is completely opposite to the conclusions, the presence of magnetic field enhances the dissociation of bound state, obtained from the inverse magnetic catalysis model in Ref.\cite{Zhao:2021ogc}. While the change of effective mass is non-trivial as is displayed in Fig.\ref{Figp2}. When magnetic field is parallel to the polarization, the effective mass reduces with the increasing magnetic field in the full temperature regime. When magnetic field is perpendicular to the polarization, in the lower and higher temperature regimes, the increasing magnetic field enlarges the effective mass, while that reduces the effective mass for the middle-temperature regime. In addition, by comparing the left and right panels, one can easily find that the suppression effect is stronger when the magnetic field is perpendicular to polarization, but the effective mass is smaller when the magnetic field is parallel to polarization which is consistent with the conclusion from inverse magnetic catalysis model in Ref.\cite{Zhao:2021ogc}.

On balance, an interesting conclusion for the $J/\Psi$ state can be obtained: Increasing temperature and chemical potential all enhance the dissociation effect, while the magnetic field suppresses the dissociation effect. The dependence of effective mass on magnetic field and chemical potential is non-trivial, which is strictly dependent on the temperature.

\subsection{Turning on angular momentum }
\subsubsection{ The case of magnetized-rotating QGP  with MC }
%
\begin{figure}[t]
  \centering
     \includegraphics[width=0.49\textwidth]{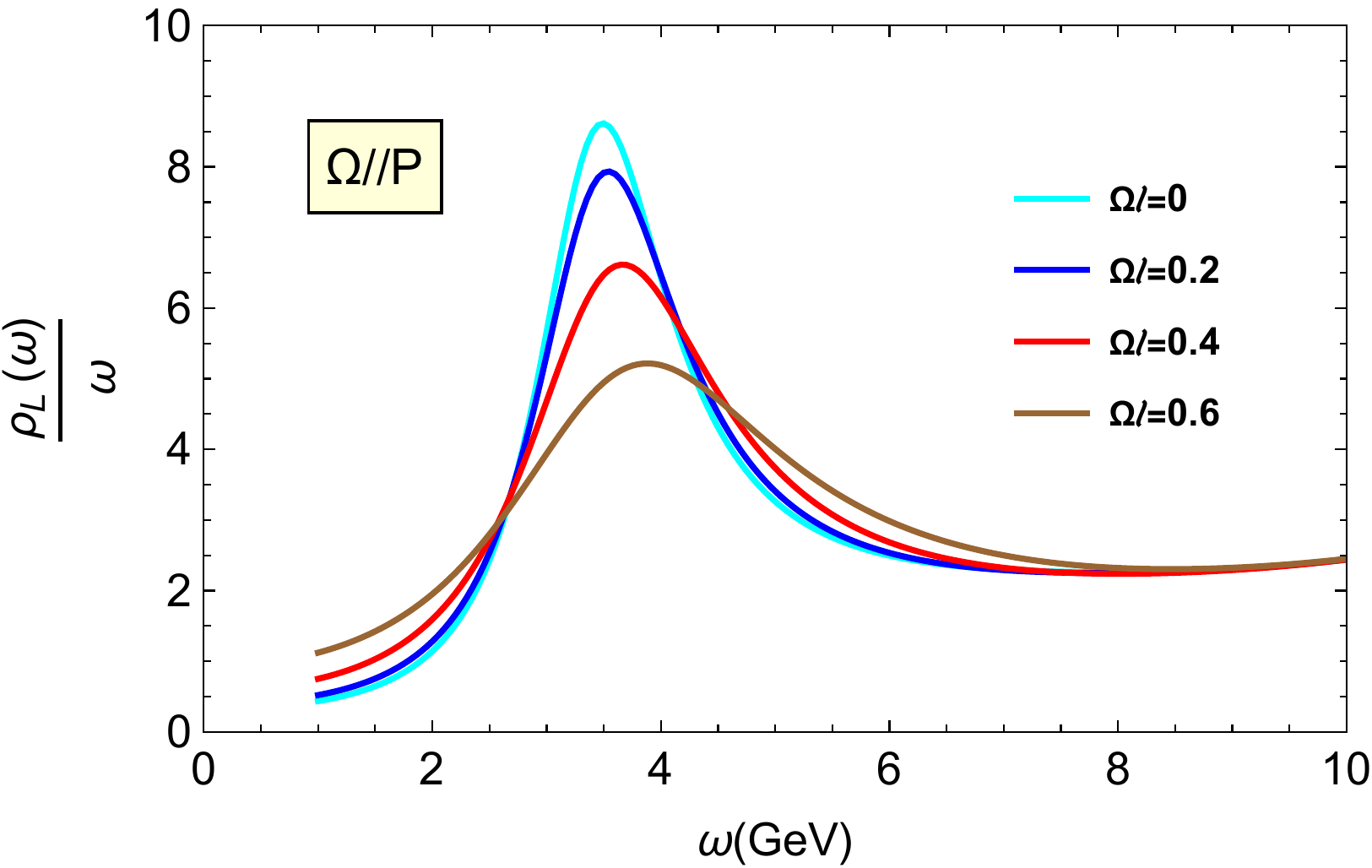}
     \includegraphics[width=0.49\textwidth]{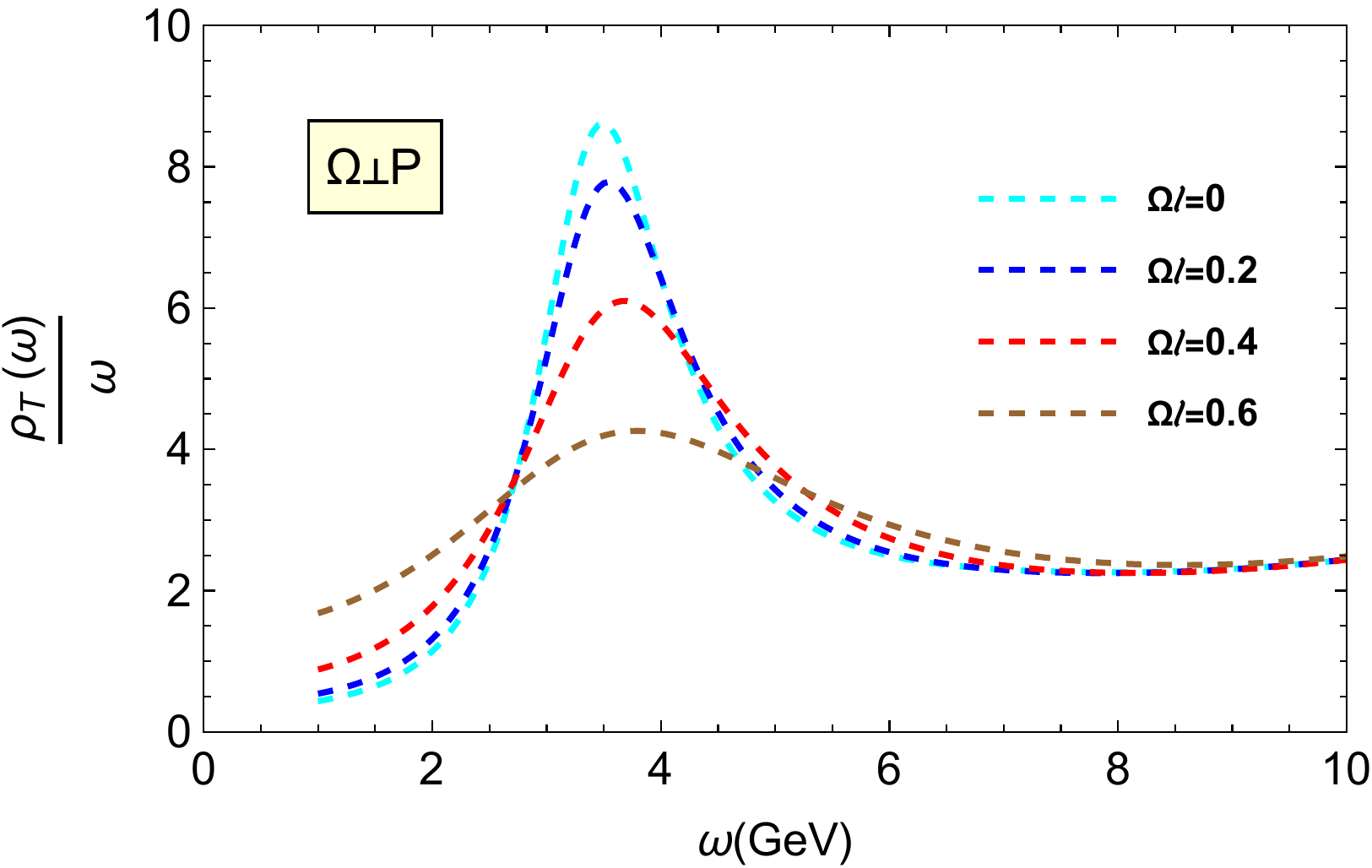}\\
  \caption{The spectral functions of the $J/\Psi$ state with different angular velocity $\Omega l$ at $\mu=0\,\text{GeV}$,$B=0\,\text{GeV}^2$ and $T=0.6\,\text{GeV}$ for magnetic catalysis model. From top to bottom, the curves represent $\Omega l=0, 0.2, 0.4, 0.6$ respectively. }\label{Fig5}
\end{figure}
\begin{figure}[t]
  \centering
     \includegraphics[width=0.49\textwidth]{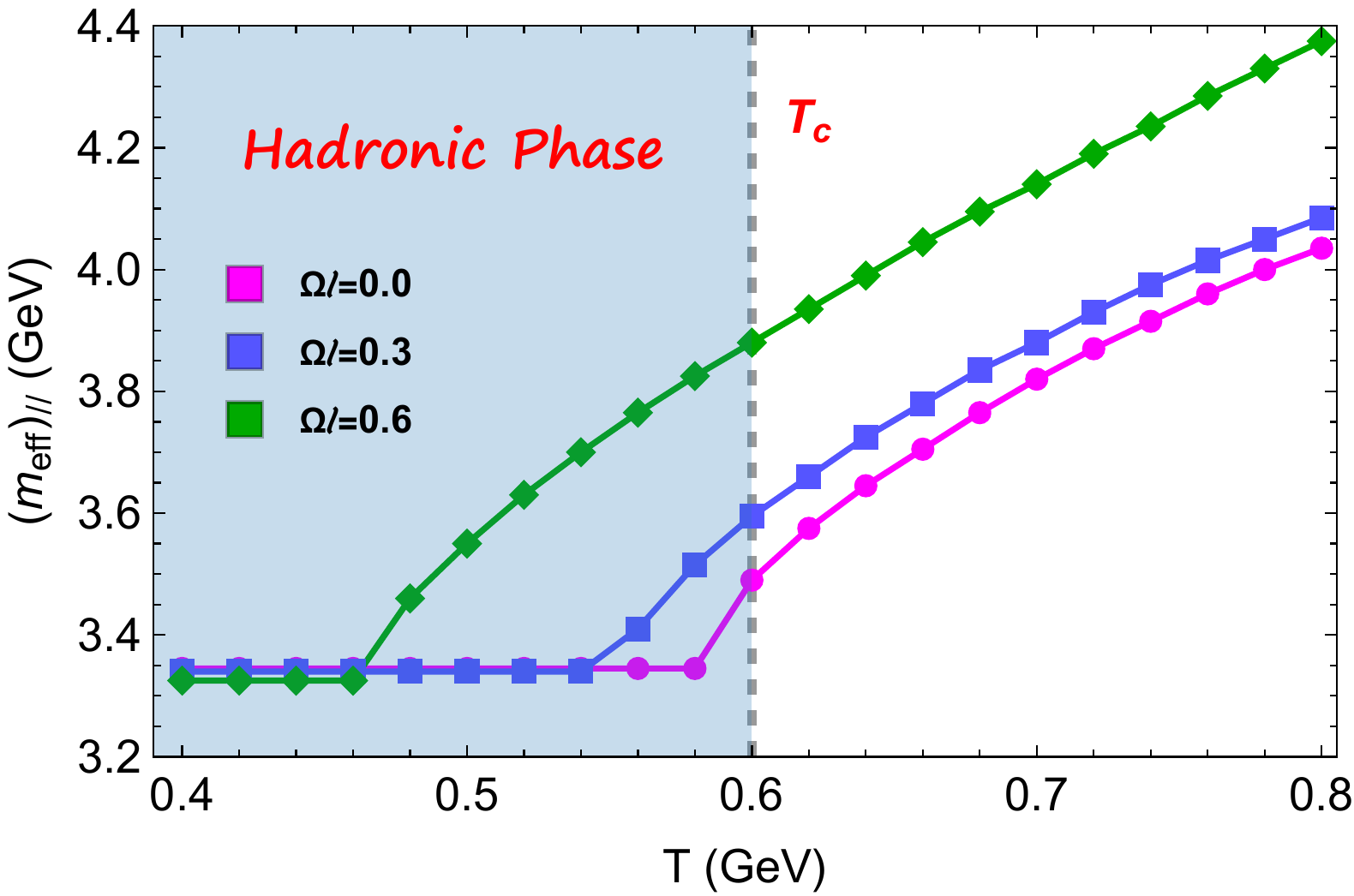}
     \includegraphics[width=0.49\textwidth]{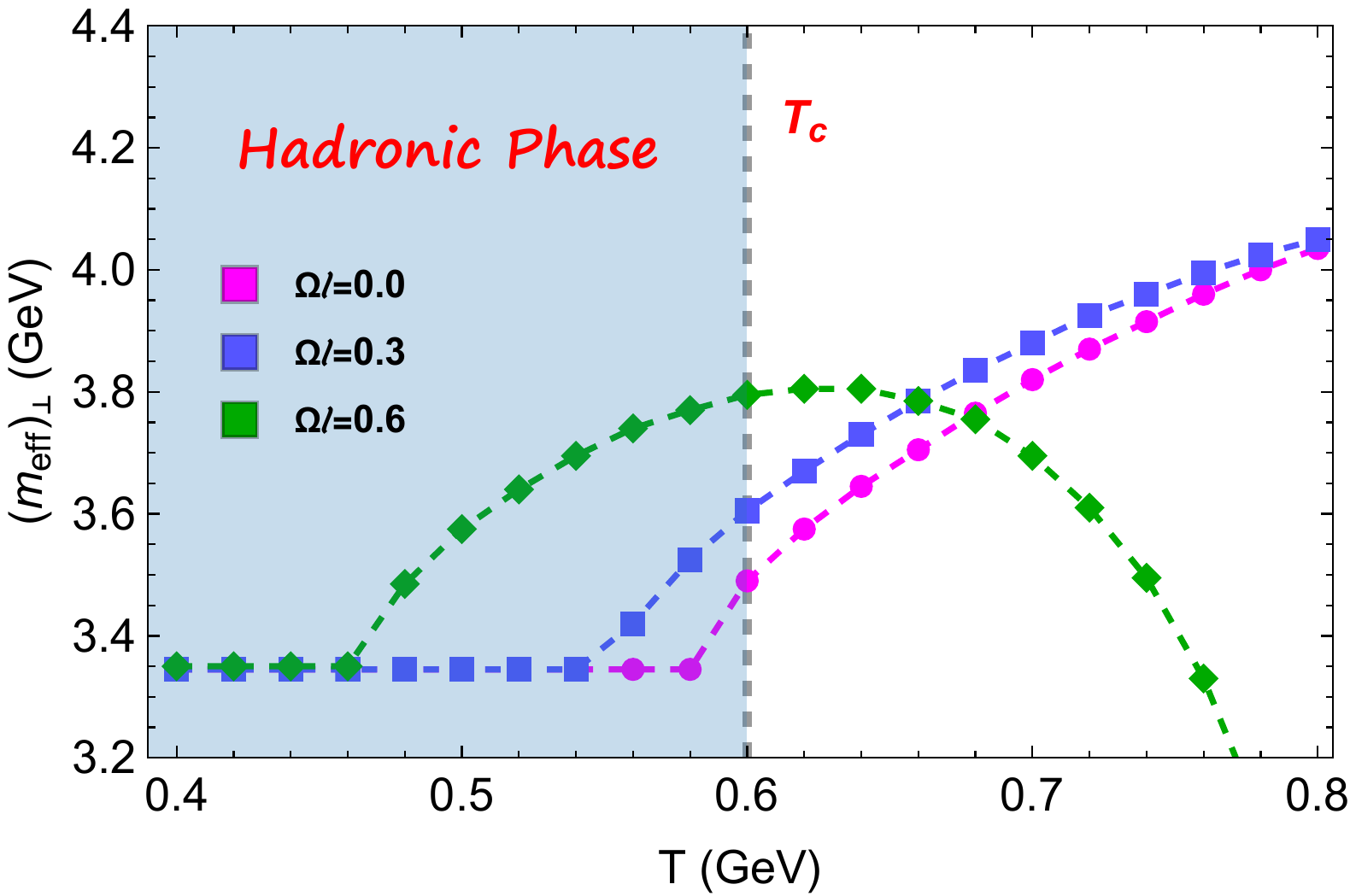}\\
  \caption{The effective mass of $J/\Psi$ state as a function of temperature in different angular velocities for magnetic catalysis model. The left picture is for angular velocity parallel to polarization and the right one is for angular velocity perpendicular to polarization. }\label{Figp3}
\end{figure}
In Fig.\ref{Fig5} we show the behavior of spectral function for different angular velocities at $\mu=0\,\text{GeV}, B=0\,\text{GeV}^2, T=0.6\,\text{GeV}$. The left figure is for the rotating direction parallel to polarization and the right figure is for the rotating direction perpendicular to polarization. Whether the rotating direction is parallel or perpendicular to the polarization, increasing angular velocity reduces the height and enlarges the width, which means angular velocity speed up the dissociation effect. In addition, by comparing the parallel and the perpendicular cases in Fig.\ref{Fig5}, it is found that the dissociation effect is stronger for the perpendicular case. The angular velocity dependence of effective mass is shown in Fig.\ref{Figp3}. The results suggest that angular velocity increases the effective mass. It is noted that the shadow region denotes the hadronic phase, so the result is untrustworthy in the regime. Because we introduce the rotation in the black hole metric on the side of gravity corresponding to the rotation of QGP on the side of the gauge field. The shadow region should be described by introducing the rotation in the thermal gas solution rather than the black hole solution. An interesting behavior can be observed in the perpendicular case when the angular velocity is larger.  The effective mass has a maximum near the phase transition temperature, which may be regarded as a dissociation signal of the $J/\Psi$ state. This point can also be checked in Fig.\ref{Fig5}, the peak is very low suggesting the quarkonium dissociation. And we find the non-trivial behavior from effective mass occurs only in the perpendicular case. This conclusion also well shows that the dissolution effect is stronger in the perpendicular case, which is consistent with the conclusion obtained by using the height of the spectral function peak in Fig.\ref{Fig5}.
\begin{figure}[t]
  \centering
  \includegraphics[width=0.6\textwidth]{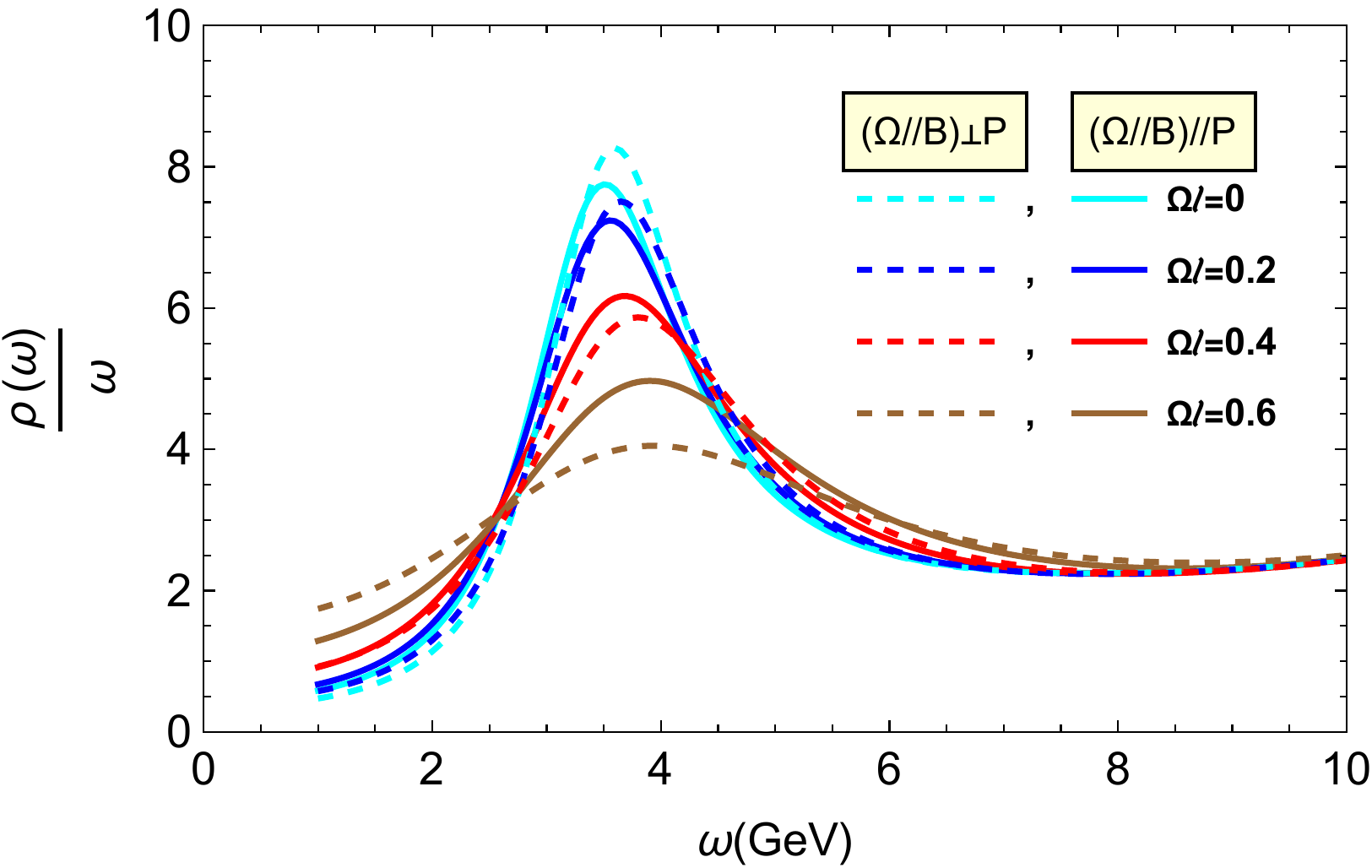}\\
  \caption{The spectral functions of the $J/\Psi$ state, by considering the superposition effect of magnetic field and rotation, with different angular velocity $\Omega l$ at $\mu=0\,\text{GeV}$, $B=0.5\,\text{GeV}^2$ and $T=0.63\,\text{GeV}$ for magnetic catalysis model.}\label{Fig8}
\end{figure}
\begin{figure}[t]
  \centering
  \includegraphics[width=0.49\textwidth]{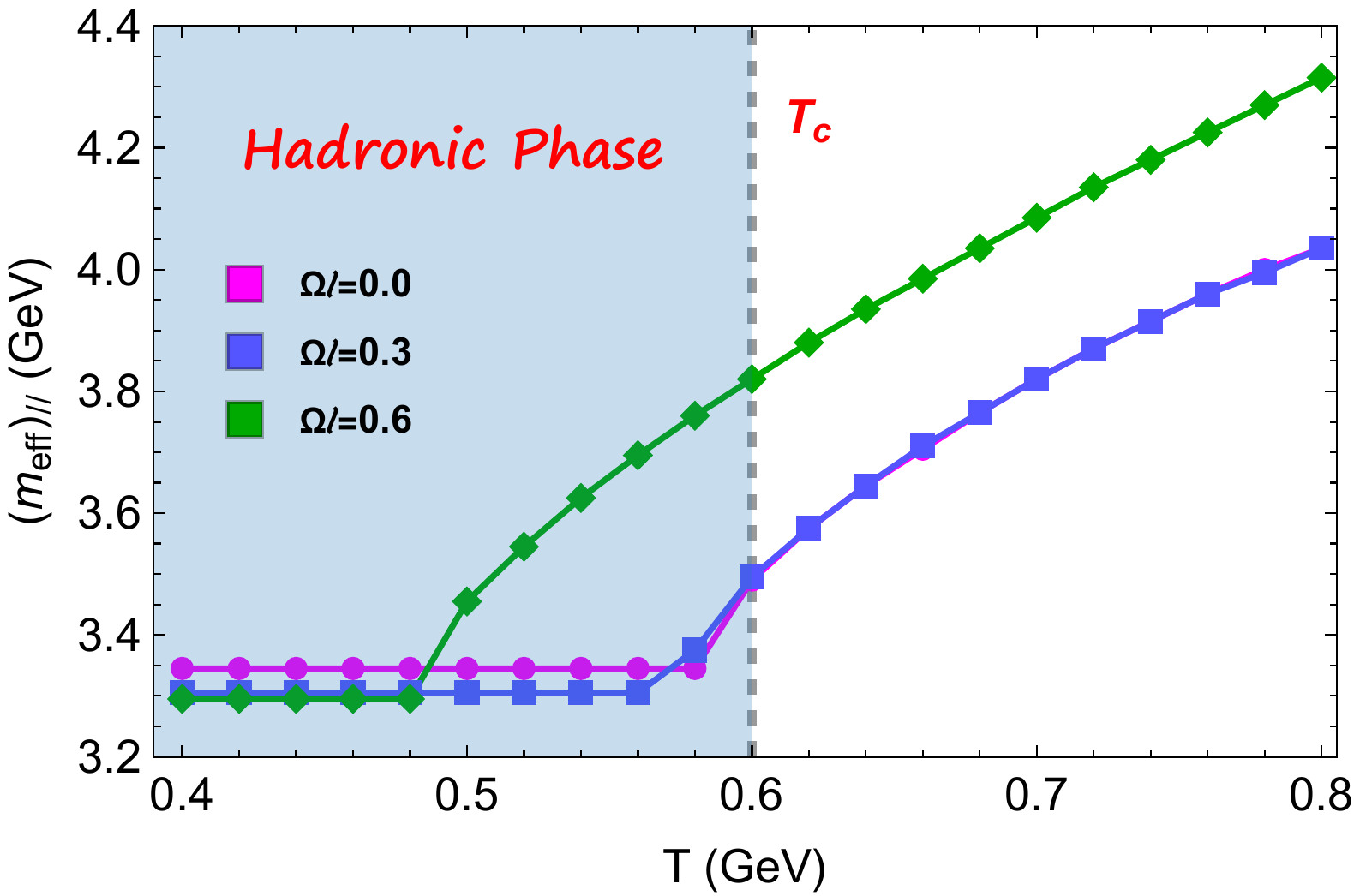}
  \includegraphics[width=0.49\textwidth]{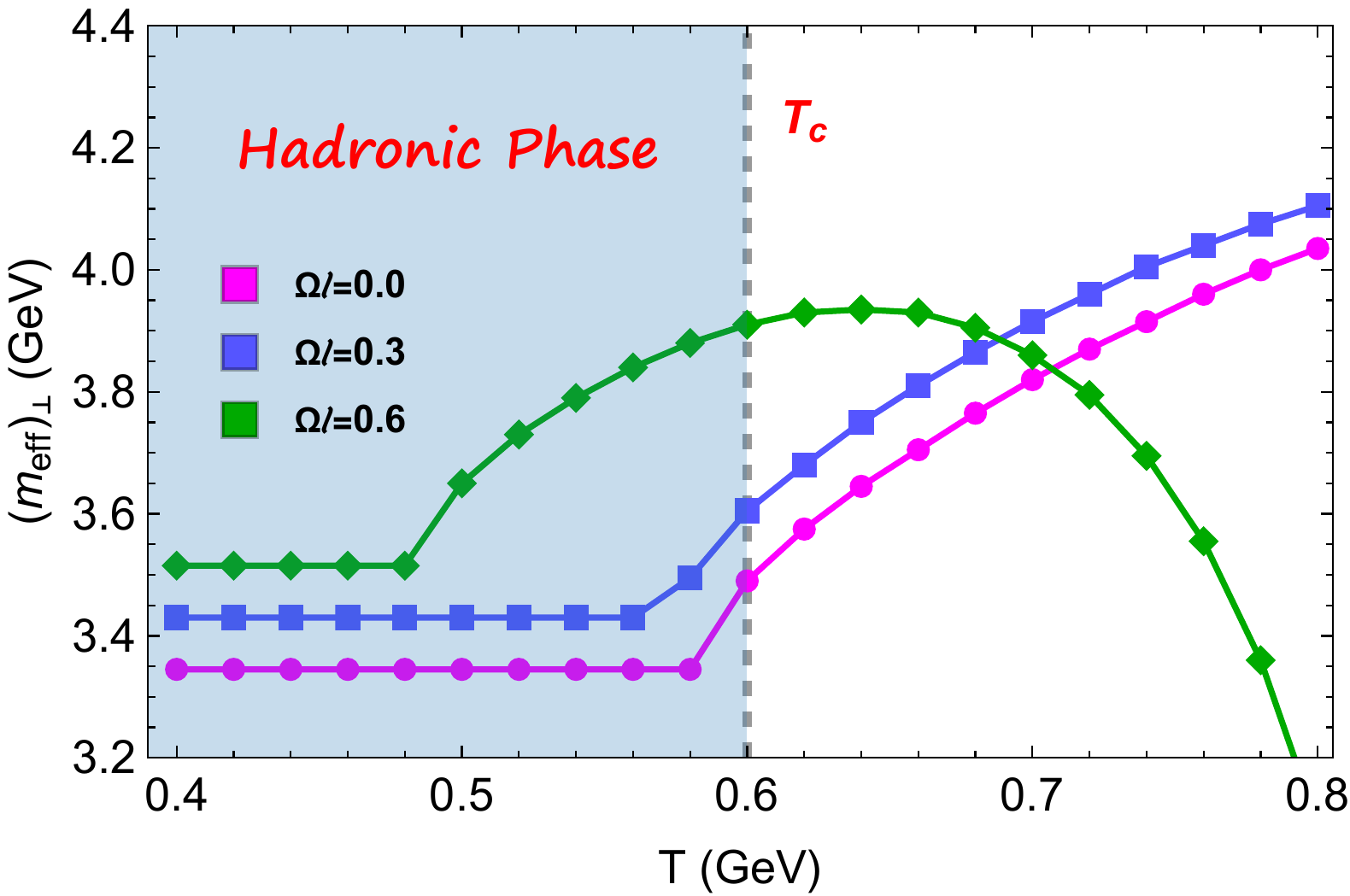}\\
  \caption{The effective mass of the $J/\Psi$ state, by considering the superposition effect of magnetic field and rotation, with different angular velocity $\Omega l$ at $\mu=0\,\text{GeV}$, $B=0.5\,\text{GeV}^2$ and $T=0.63\,\text{GeV}$ for magnetic catalysis model.}\label{Figp4}
\end{figure}
To sum up, we find that temperature, chemical potential, and angular velocity promote the dissolution effect, while magnetic field suppresses it. So one can conclude that there must be a competition effect between magnetic field and temperature, chemical potential, and angular velocity. As an example, we show the behavior of spectral function for the superposition effect of rotation and magnetic field ( here we only show the case of magnetic field parallel to rotating direction ) in Fig. \ref{Fig8}. One can find that there is a critical angular velocity $\Omega_{crit}(B)l\sim 0.3$ which is the result of the competition between the magnetic field effect and the rotating effect. Here $\Omega_{crit}(B)$ denotes that the value of critical angular velocity $\Omega_{crit}$ depends on the size of magnetic field $B$. When $\Omega<\Omega_{crit}(B)$, the magnetic field effect plays a leading role and suppresses the dissociation effect, which leads to the dissociation effect being stronger when the rotating direction (magnetic field) is parallel to the polarization. However, when $\Omega>\Omega_{crit}(B)$, the rotating effect is dominant, which causes the dissolution effect to be more intense when the rotating direction is perpendicular to the polarization. In Fig.\ref{Figp4}, we show the influence of the superposition effect between magnetic field and rotation on effective mass. One can find, for the parallel case, the magnetic field suppresses the growth of effective mass caused by rotation, while the magnetic field speeds up the growth of effective mass for the perpendicular case. But for this intensity of magnetic field, the behavior of effective mass still is dominant by rotation in the QGP phase. Although the result from the hadronic phase is untrustworthy, we can use it as a reference to conclude the effective mass in hadronic phase is dominant by an intense magnetic field instead of rotation. We will check this conclusion in the future.
\subsubsection{The case of magnetized-rotating QGP  with  IMC }
%
\begin{figure}[t]
  \centering
     \includegraphics[width=0.49\textwidth]{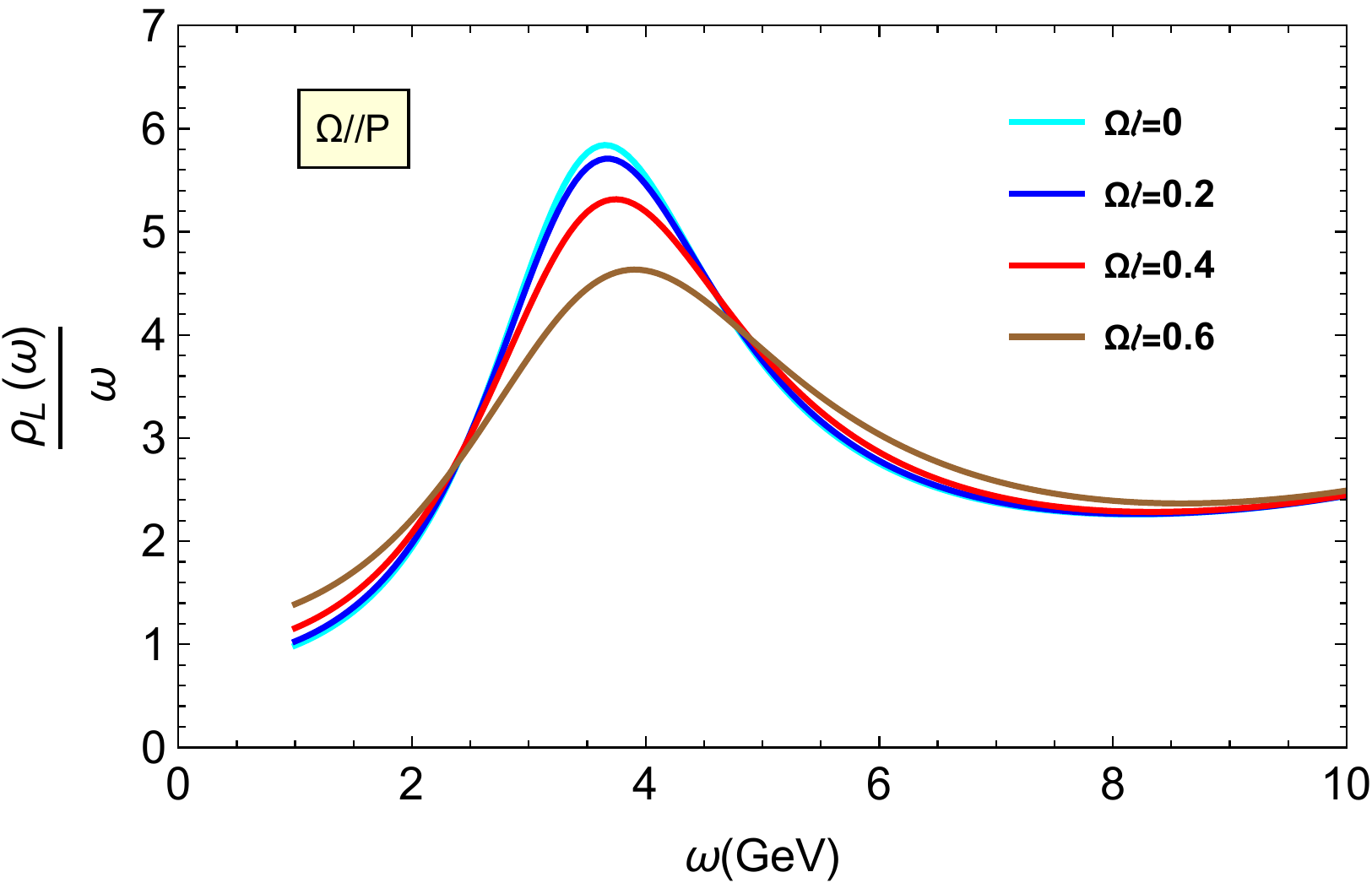}
     \includegraphics[width=0.49\textwidth]{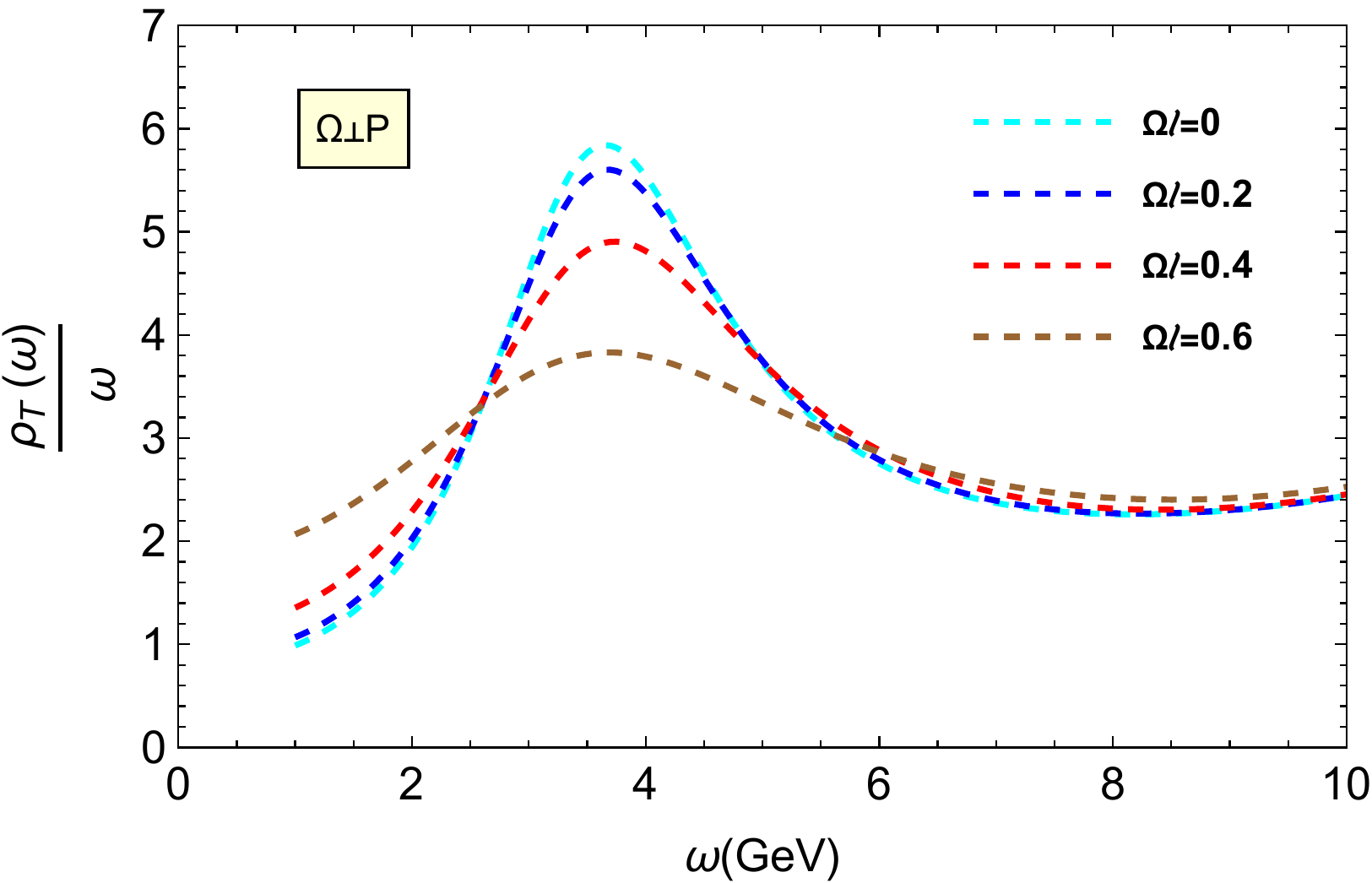}\\
  \caption{The spectral functions of the $J/\Psi$ state with different angular velocity $\Omega l$ at $\mu=0\,\text{GeV}$,$B=0\,\text{GeV}$ and $T=0.6\,\text{GeV}$ for inverse magnetic catalysis model. From top to bottom, the curves represent $\Omega l=0, 0.2, 0.4, 0.6$ respectively. }\label{Fig17}
\end{figure}
\begin{figure}[t]
  \centering
  \includegraphics[width=0.6\textwidth]{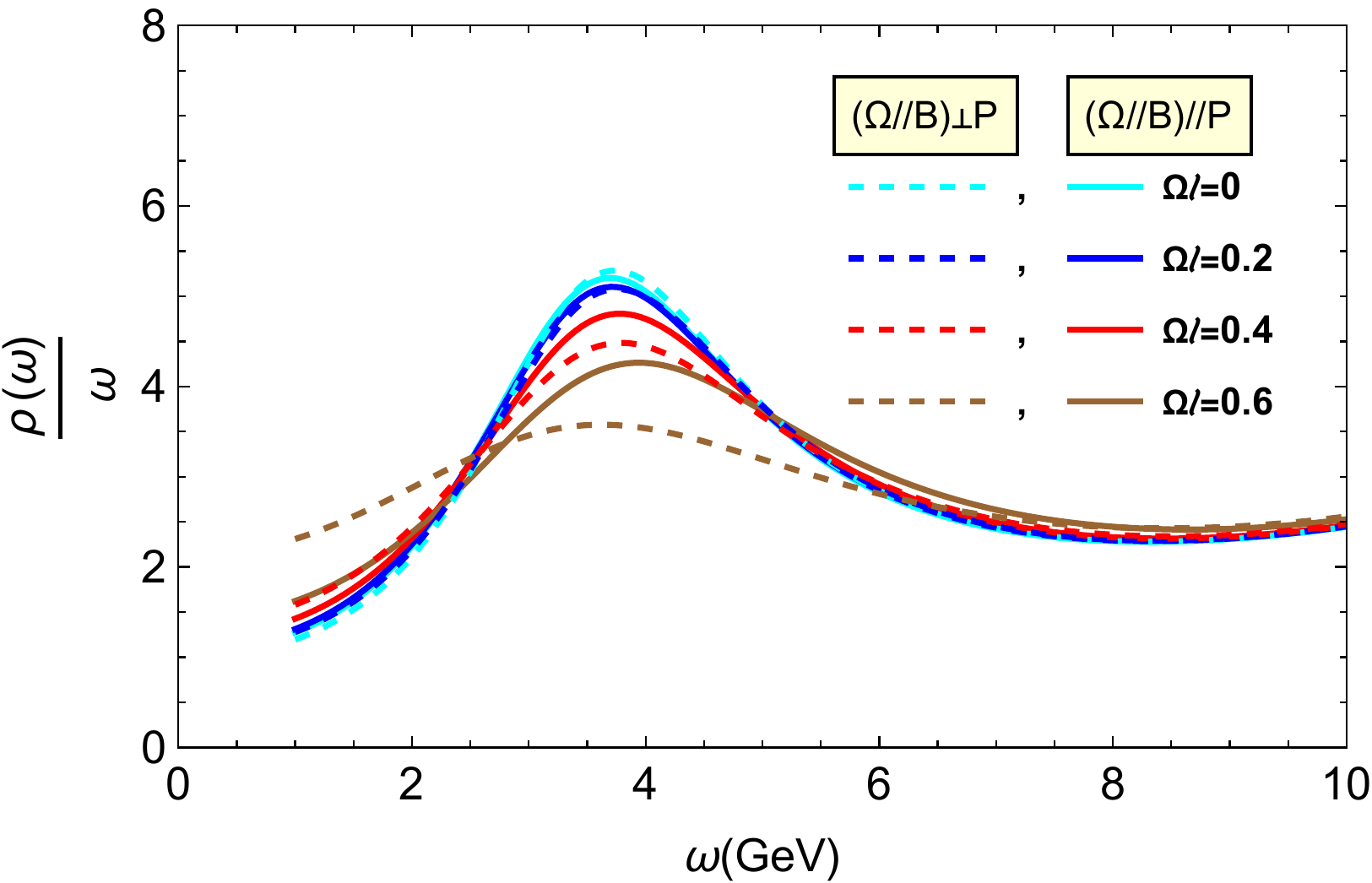}\\
  \caption{The spectral functions of the $J/\Psi$ state, by considering the superposition effect of magnetic field and rotation, with different angular velocity $\Omega l$ at $\mu=0\,\text{GeV}$,$B=0.5\,\text{GeV}$ and $T=0.63\,\text{GeV}$ for inverse magnetic catalysis model.}\label{Fig20}
\end{figure}
In addition, as a parallel study, we introduce the rotation in a holographic inverse magnetic catalysis model(refer to appendix\ref{App2} for more details). As we focus primarily on the impact of rotation on the dissociation effect of $J/\Psi$ in the two different magnetic field models, we present only the behavior of the spectral function in Fig.\ref{Fig17} and Fig.\ref{Fig20}. From Fig.\ref{Fig17}, one can find that growing angular velocity promotes the dissociation effect and the melting effect is stronger for the direction of angular velocity perpendicular to the direction of polarization. In addition, by considering the superposition effect of magnetic field and rotation, for a smaller angular velocity, Fig.\ref{Fig20} straightforwardly illustrates that the dissociation effect is stronger for the direction of angular velocity parallel to the direction of polarization, which indicates the dissociation effect is dominant by magnetic field. As pointed out in Ref.\cite{Zhao:2021ogc}, the melting effect produced by the magnetic field is stronger in the parallel case. For a larger angular velocity, an opposing conclusion is obtained that suggests the dissociation effect is controlled by angular velocity. Although the magnetic field exhibits distinctly different behaviors in these two models, the effect of rotation on the dissociation effect of $J/\Psi$ is similar.

In a word, the influences of rotation on the dissociation effect of the bound state can be summarized as follows: (1) As the increase of angular velocity, the dissociation effect becomes stronger and the effective mass becomes larger and the dissociation effect is stronger for the perpendicular case. (2)The superposition effect of magnetic field and rotation,
\begin{figure}[t]
  \centering
  \includegraphics[width=0.49\textwidth]{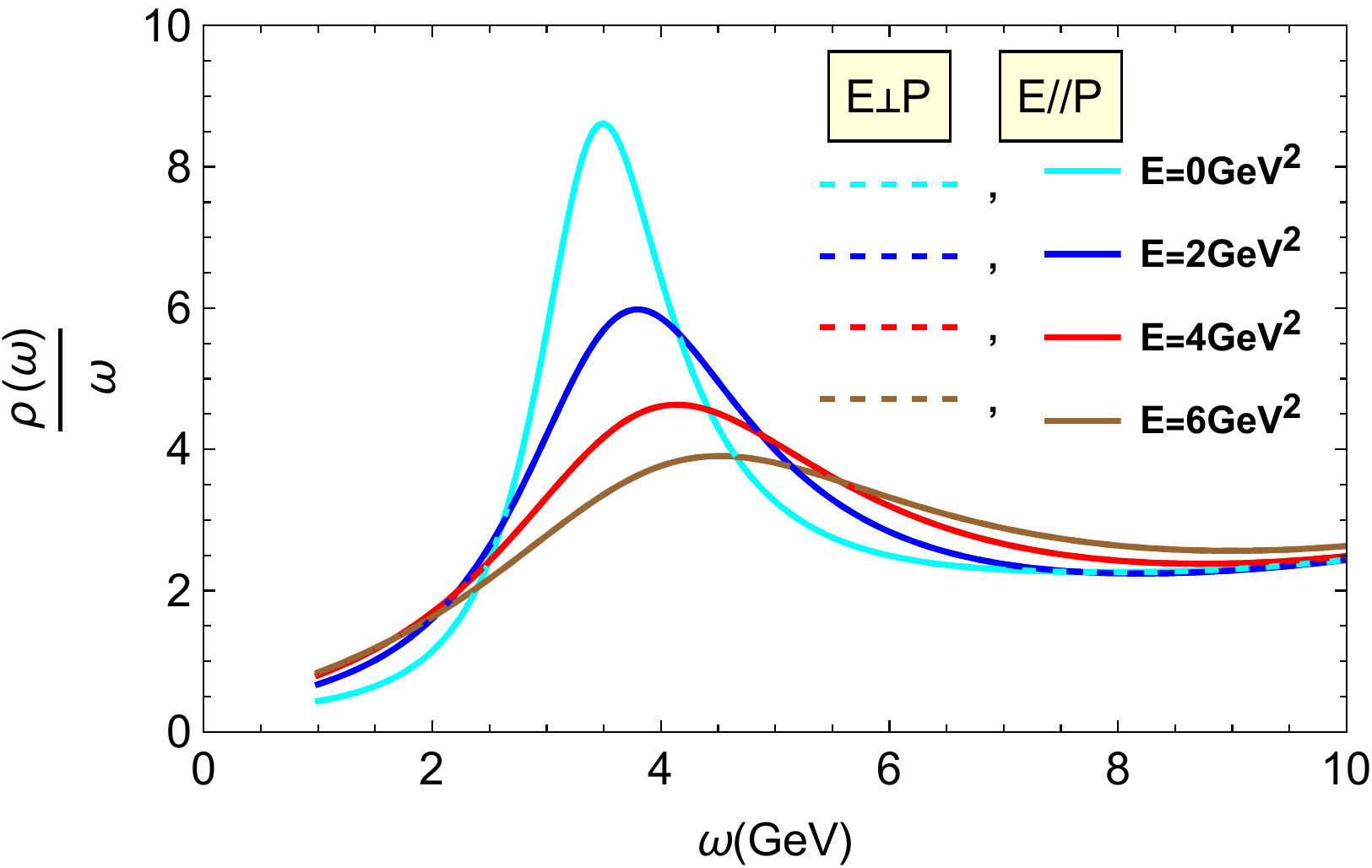}
  \includegraphics[width=0.47\textwidth]{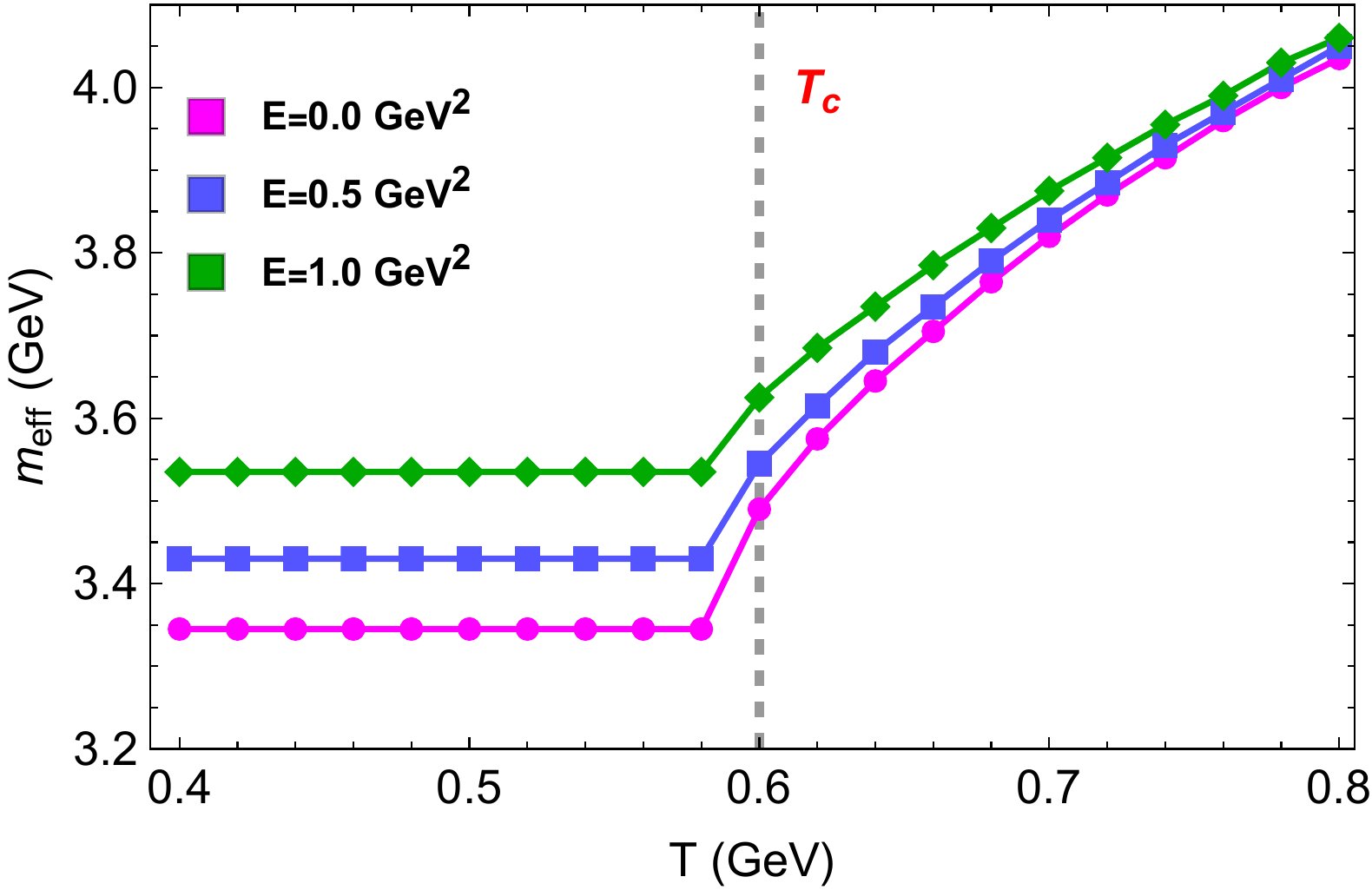}\\
  \caption{Left: The spectral functions of the $J/\Psi$ state with different electric field $E$ at $\mu=0\,\text{GeV}$,$B=0\,\text{GeV}^2$ and $T=0.6\,\text{GeV}$ for magnetic catalysis model. Right: The effective mass of $J/\Psi$ state as a function of temperature in different electric fields for magnetic catalysis model. }\label{Fig12}
\end{figure}
the strength of the dissolution effect depends entirely on the interplay between the magnetic field and the rotation effect and the behavior of effective mass is non-trivial. (3)Whether the magnetic field behaves as magnetic catalysis or inverse magnetic catalysis, the conclusion from the rotating effect is similar.
\subsection{Turning on electric field}
%
\begin{figure}[t]
  \centering
  \includegraphics[width=0.6\textwidth]{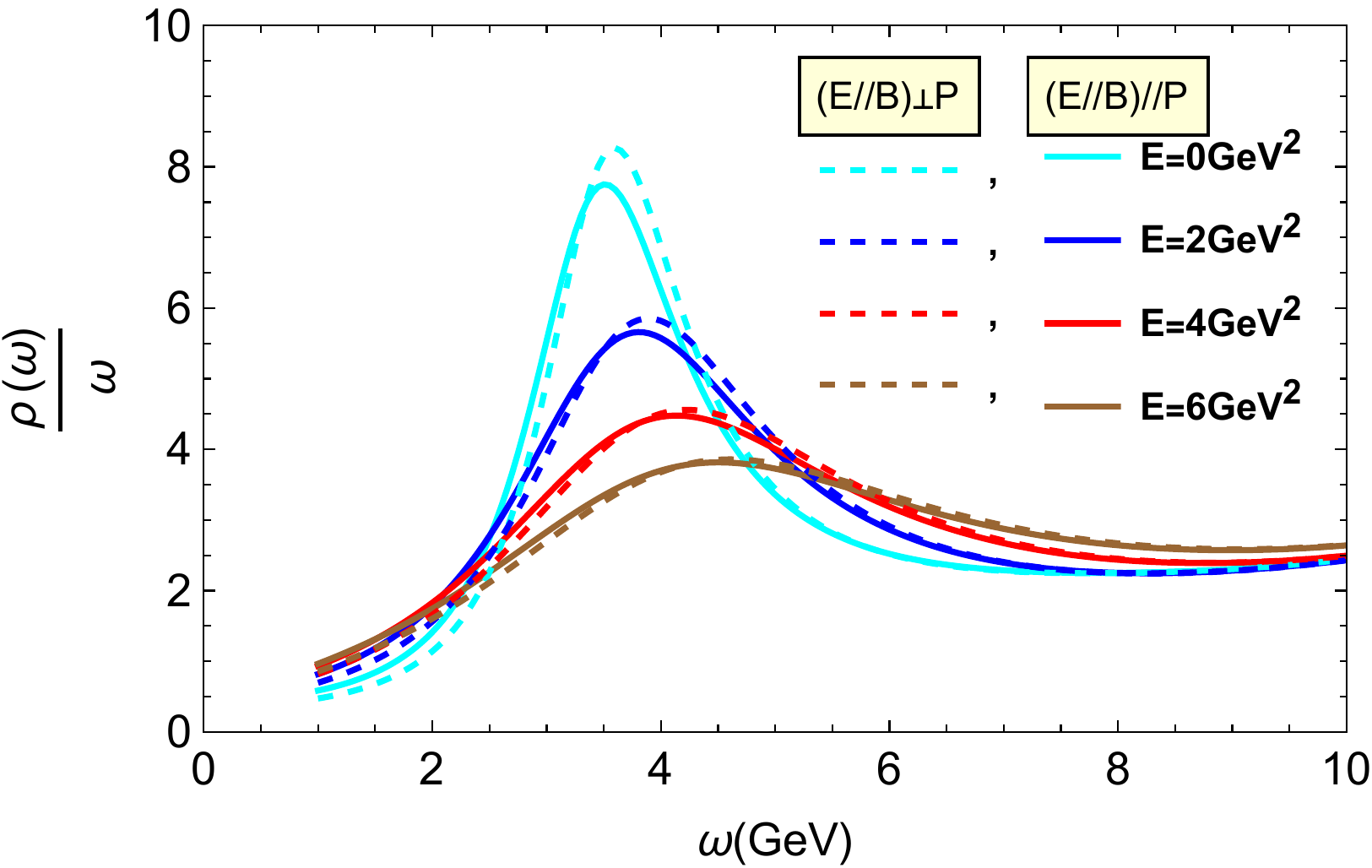}\\
  \caption{The spectral functions of the $J/\Psi$ state with different electric field $E$ at $\mu=0\,\text{GeV}$, $B=0.5\,\text{GeV}^2$ and $T=0.63\,\text{GeV}$ for magnetic catalysis model.}\label{Fig14}
\end{figure}
\begin{figure}[t]
  \centering
  \includegraphics[width=0.49\textwidth]{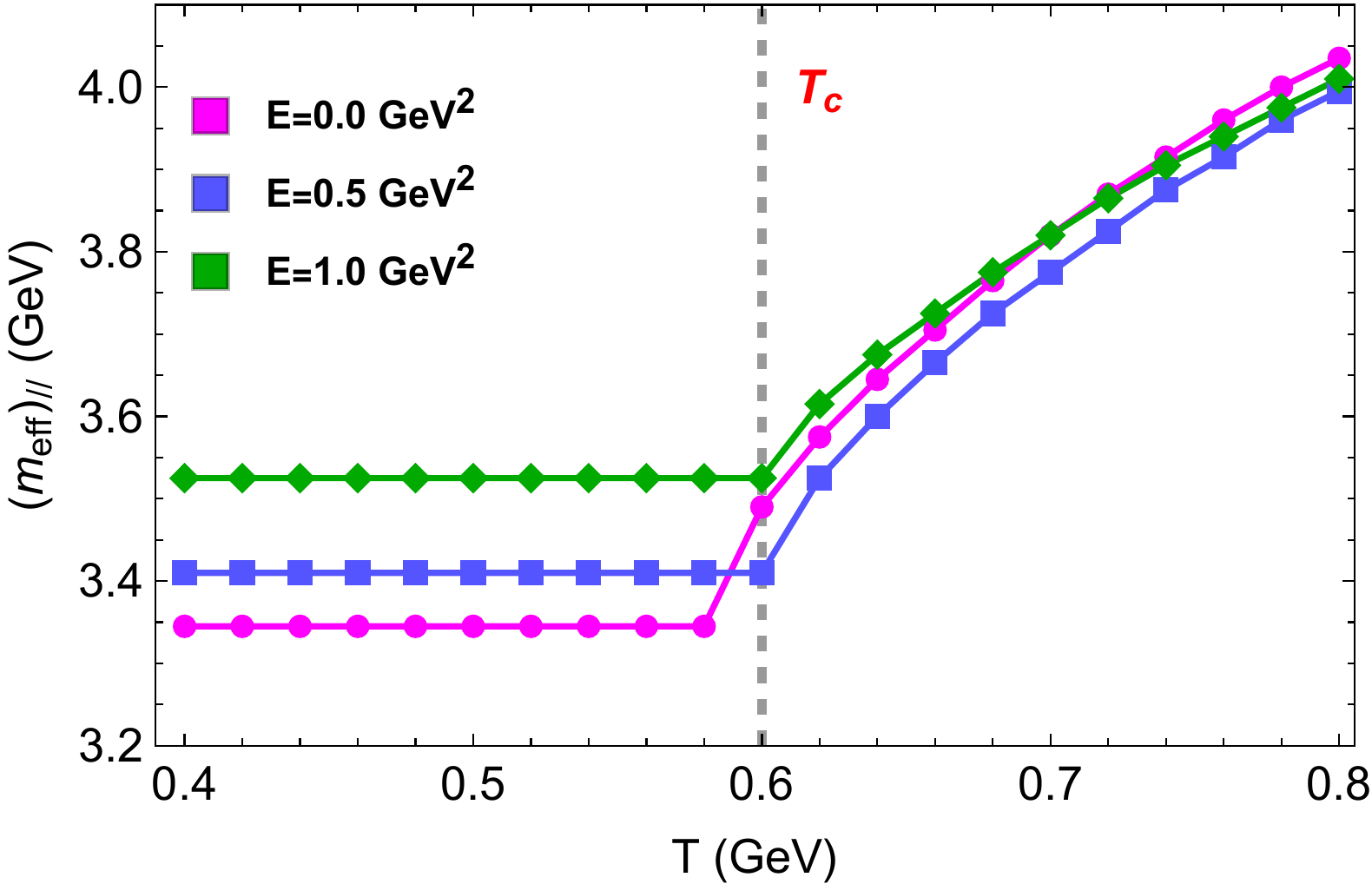}
  \includegraphics[width=0.49\textwidth]{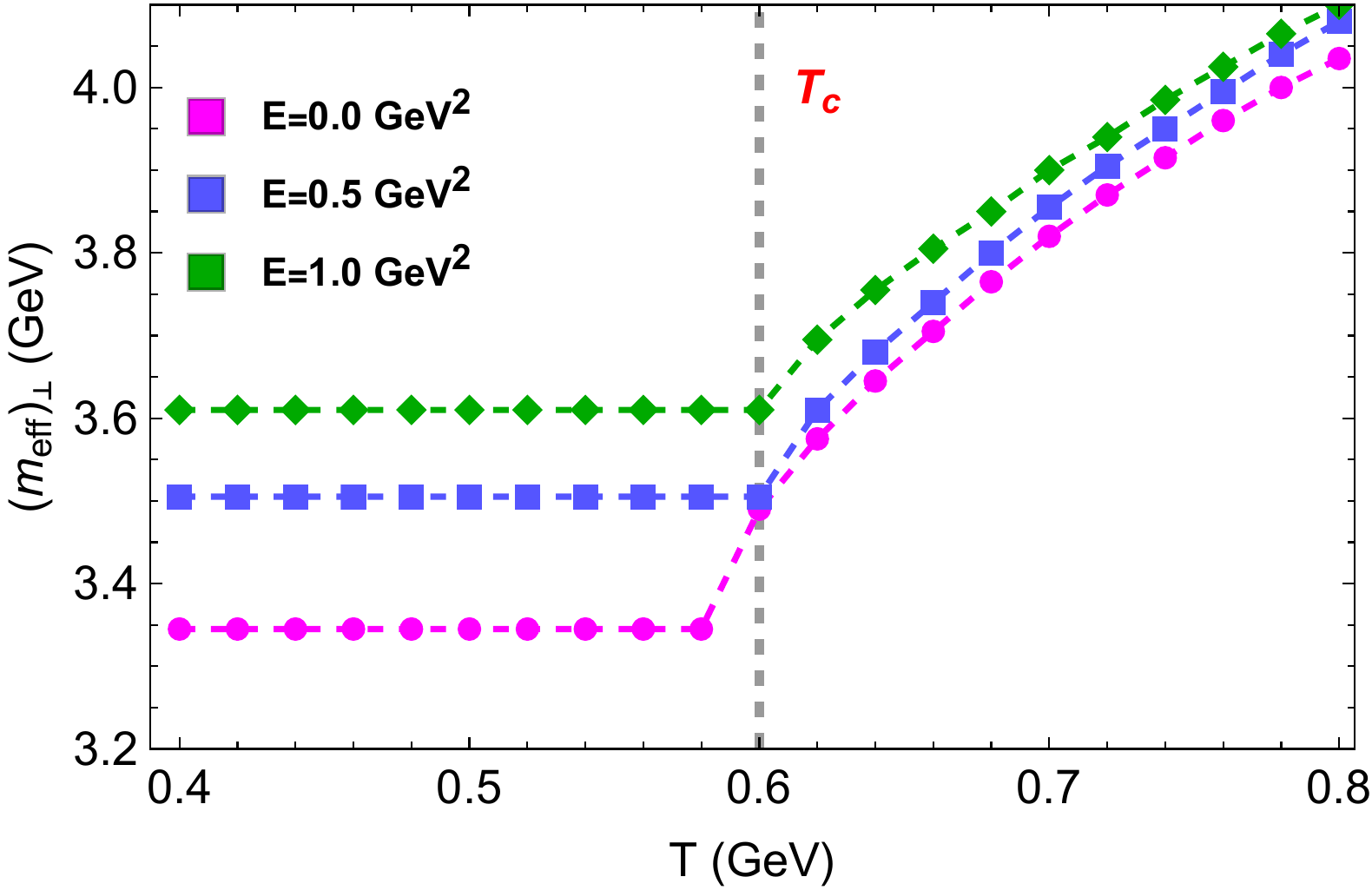}
  \\
  \caption{The effective mass of the $J/\Psi$ state, by considering the superposition effect of magnetic field and electric field, with different electric field $E$ at $\mu=0\,\text{GeV}$, $B=0.5\,\text{GeV}^2$ and $T=0.63\,\text{GeV}$ for magnetic catalysis model.}\label{Figp5}
\end{figure}
In Fig.\ref{Fig12}, we draw the spectral function and effective mass of $J/\Psi$ state for varying  electric field at $B=0\,\text{GeV}^2, \mu=0 \,\text{GeV}, T=0.6\,\text{GeV}$. The result shows that the dissociation effect and the effective mass are enhanced by the increasing electric field. But one can find that, whether E-field is parallel or perpendicular to polarization, the dissociation effect is the same.

Since the electric field and magnetic field have opposite interactions on the dissociation effect and effective mass. Next, we take into account the superposition of the magnetic field and electric field. As an example, we plot the spectral function and effective mass with respect to the electric field for the electric field parallel to the magnetic field in Fig.\ref{Fig14} and Fig.\ref{Figp5}, respectively. One can find that the difference from the dissociation effect between the magnetic field parallel and perpendicular to polarization reduces with the increase of electric field and the difference will vanish when the electric field is large enough. The effective mass in the lower temperature regime entirely depends on the electric field, however, in the higher temperature regime, that is determined by the interplay between the magnetic field and electric field.

To make a long story short, the influence of electric field on spectral function can be summarized as follows: (1) The electric field  enhances the dissociation effect and enlarges the effective mass of the bound state $J/\Psi$. (2) The dissociation effect is the same for the electric field parallel and perpendicular to polarization. (3) Increasing electric field reduces the difference from the dissociation effect between the magnetic field parallel and perpendicular to polarization and the difference will vanish for a sufficiently large electric field.

\section{Summary and discussion }\label{sec:05}

In this paper, by calculating the spectral function described by Eq.\eqref{eq20}, we investigate the dissociation effect of bound state $J/\Psi$ in a holographic magnetic catalysis model. In order to more truly simulate the non-central collision environment of extremely relativistic heavy ions, we consider a total of five effects, thermal effect, density effect, magnetic effect, rotation effect, and electric effect.

The results show that both temperature and chemical potential promote the dissociation effect and enlarge the effective mass of heavy quarkonium $J/\Psi$ in the QGP phase, while the magnetic field suppresses the dissociation effect and the behavior of effective mass is non-trivial. Interestingly, increasing magnetic field reduces the effective mass for the parallel case but enlarges the effective mass in the lower and higher temperature regimes for the perpendicular case. In addition, by considering the rotating QGP background, we obtain that the dissociation effect becomes stronger and the effective mass becomes larger with the increase of angular velocity.

In view of the completely opposite behavior of the spectral functions generated by magnetic field and rotation, we consider the superposition effect of magnetic field and angular velocity. As an example, we show the behavior of spectral function in the case of a magnetic field parallel to the rotating direction. It is found that there exists a critical angular velocity $\Omega_{crit}(B)$ depending on the magnetic field. For $\Omega<\Omega_{crit}(B)$,  the magnetic field plays a leading role and suppresses the dissociation effect, which leads to the dissociation effect being stronger and the effective mass being smaller when the rotating direction (magnetic field) is parallel to the polarization. For $\Omega>\Omega_{crit}(B)$, the rotating effect is dominant, which causes the dissolution effect to be more intense and the effective mass is larger when the rotating direction is perpendicular to the polarization. As a parallel study, we also examine the rotation effect in the holographic inverse magnetic catalysis model, although the magnetic field exhibits distinctly different behaviors in these two models, the impact of rotation on the dissociation effect of $J/\Psi$ is similar.

Besides, we consider the influence of electric field on spectral functions, the calculation indicates the increasing electric field enhances the dissociation effect and enlarges the effective mass. It is noted that whether the electric field is parallel or perpendicular to polarization, the dissociation effect is the same. In addition, we find that increasing electric field could reduce the difference in spectral functions between magnetic field parallel and  perpendicular to polarization until the difference vanishes.

According to the interesting conclusion of this paper, in the future,  it will be desirable to study the magneto-rotational and electric-rotational dissociation effects of heavy mesons and analyze the competition between the two effects. Of course, one can also study other heavy vector mesons, such as $\Upsilon(1S)$, or other physical quantities, such as configuration entropy and QNMs~\cite{Zhao:2023yry}, the spin density matrix element $\rho_{00}$~\cite{Sheng:2022wsy, Sheng:2022ssp}. As paper\cite{Zhao:2021ogc} points out, the properties of $J/\Psi$ and $\Upsilon(1S)$ are very different. In addition, one can also consider the influence of differential rotation on the dissociation effect. Because the rotation generated in the process of heavy ion collision depends on the distance to the rotating axis, which implies that angular velocity has an attenuation along the rotating radius,  it is very necessary to consider a differential rotation to simulate the RHIC and LHC experiments. To establish a connection between holographic theory and experimental observations, one can consult Ref.\cite{Li:2023mpv} for the calculation of collision energy $\sqrt{S_{NN}}$.

\section*{Acknowledgements}
We would like to thank Song He and Hai-cang Ren for the useful discussions. This work is supported in part by the National Key Research and Development Program of China under Contract No. 2022YFA1604900. This work is also partly supported by the National Natural Science Foundation of China (NSFC) under Grants No. 12275104, No. 11890711, No. 11890710, and No. 11735007.

\appendix
\section{The effect of rotating QGP on $J/\Psi$ dissociation in a holographic inverse magnetic catalysis model}\label{App2}

Here, we consider a 5d EMD gravity system whose Lagrangian is given as
\begin{equation}\label{B1}
  \mathcal{L} =\sqrt{-g}(R-\frac{g_1(\phi_0)}{4}F_{(1)\mu\nu}F^{\mu\nu}-\frac{g_2(\phi_0)}{4}F_{(2)\mu\nu}F^{\mu\nu}
  -\frac{1}{2}\partial_\mu\phi_0\partial^\mu\phi_0-V(\phi_0)).
\end{equation}
where $F_{(i)\mu\nu}$ ($i=1,2$) is the field strength tensor for U(1) gauge field, $g_i(\phi_0)$ ($i=1,2$) denotes the gauge coupling kinetic function, $\phi_0$ represents the dilaton field, $V(\phi_0)$ is the potential of the $\phi_0$ (see\cite{Bohra:2019ebj} for exact expression). By introducing an external magnetic field in $x_1$ direction,  the metric ansatz in Einstein frame can be written as ~\cite{Bohra:2019ebj}
\begin{align}\label{B2}
 ds^2 &= \frac{R^2S(z)}{z^2}(-f(z)dt^2+dx_1^2+e^{B^2z^2}(dx_2^2+dx_3^2)+\frac{dz^2}{f(z)}), \notag \\
\phi_0 & =\phi_0(z),\quad A_{(1)\mu}=A_t(z)\delta_\mu^t,\quad F_{(2)}=B dx_2\wedge dx_3,
\end{align}
with
\begin{align}\label{B3}
 f(z) &  =1+\int_0^zd\xi \xi^3e^{-B^2\xi^2-3A(\xi)}[K+\frac{\widetilde{\mu}^2}{2R_{gg}R^2}e^{R_{gg}\xi^2}],\notag \\
  K & = -\frac{1+\frac{\widetilde{\mu}^2}{2R_{gg}R^2}\int_0^{z_h}d\xi \xi^3e^{-B^2\xi^2-3A(\xi)+R_{gg}\xi^2}}{\int_0^{z_h}d\xi \xi^3e^{-B^2\xi^2-3A(\xi)}},\notag\\
  \widetilde{\mu} &= \frac{\mu}{\int_0^{z_h}d\xi\frac{\xi e^{-B^2\xi^2}}{g_1(\xi)\sqrt{S(\xi)}}},
\end{align}
where $R$ is the AdS radius, $S(z)$ labels the scale factor, $f(z)$ is the blackening function and $\mu$ denotes the chemical potential. The asymptotic boundary is at $z=0$ and $z=z_h$ denotes the location of the horizon where $f(z_h)=0$.  The concrete form of gauge coupling function $g_1$ can be determined by fitting the vector meson mass spectrum. The linear Regge trajectories for $B=0$ can be restored when
\begin{equation}\label{B4}
  g_1(z)=\frac{e^{-R_{gg}z^2-B^2z^2}}{\sqrt{S(z)}}.
\end{equation}
Here, this $B$, in units $\text{GeV}$, is the 5d magnetic field. The 4d physical magnetic field is $e\mathcal{B}\sim \frac{const}{R}\times B$ where $const=1.6$ (See details in Ref.\cite{Dudal:2015wfn}). By taking $S(z)=e^{2A(z)}$, one can obtain $R_{gg}=1.16\,\text{GeV}^2$ for heavy meson state $J/\psi$. In the following calculation, we take $A(z)=-az^2$ where $a=0.15\,\text{GeV}^2$ matching with the lattice QCD deconfinement temperature at $B=0\,\text{GeV}$ ~\cite{Dudal:2017max}.

The Hawking temperature has the following form,
\begin{equation}\label{B5}
  T(z_h,\mu,B)=\frac{-z_h^3e^{-3A(z_h)-B^2z_h^2}}{4\pi}(K+\frac{\widetilde{\mu}^2}{2R_{gg}R^2}e^{R_{gg}z_h^2}).
\end{equation}
The paper~\cite{Bohra:2019ebj} assumes that the dilaton field $\phi$ remains real everywhere in the bulk, which leads to magnetic field $B\leq B_c\simeq0.61\,\text{GeV}$. In this holographic inverse magnetic catalysis model, the deconfinement temperature is $T_c=0.268\,\text{GeV}$ at zero chemical potential and magnetic field.

In the rotating background, the metric uses Eq.\eqref{eq23}, and the temperature and chemical potential takes Eq.\eqref{eq26}. The flow equation takes Eq.\eqref{eq18} and Eq.\eqref{eq3a20}. In this holographic inverse magnetic catalysis model, we have $\tilde{g}_{tt}=\frac{f(z) S(z)}{z^2}, \tilde{g}_{x_1x_1}=\frac{S(z)}{z^2}, \tilde{g}_{x_2x_2}=g_{x_3x_3}=\frac{e^{B^2 z^2} S(z)}{z^2}, \tilde{g}_{zz}=\frac{S(z)}{z^2 f(z)}$. Finally, the spectral function can be obtained. We show the behavior of spectral functions for different angular velocities in Fig.\ref{Fig17}. In Fig.\ref{Fig20}, the influence of the superposition of angular velocity and magnetic field on the spectral function is studied. One can find the conclusion is similar to the holographic magnetic catalysis model.

	
	
	
\bibliographystyle{utphys}
\bibliography{ref}
	
\end{document}